\documentclass[%
 aip,
 amsmath,amssymb,
 reprint,%
]{revtex4-1}

\usepackage{graphicx}% Include figure files
\usepackage{dcolumn}% Align table columns on decimal point
\usepackage{bm}% bold math
\usepackage{xcolor}

\usepackage[utf8]{inputenc}
\usepackage[T1]{fontenc}
\usepackage{mathptmx}
\usepackage{etoolbox}
\usepackage{amsmath}
\usepackage{upgreek}
\usepackage{subcaption}
\usepackage{hyperref}

\makeatletter
\def\@email#1#2{%
 \endgroup
 \patchcmd{\titleblock@produce}
  {\frontmatter@RRAPformat}
  {\frontmatter@RRAPformat{\produce@RRAP{*#1\href{mailto:#2}{#2}}}\frontmatter@RRAPformat}
  {}{}
}%
\makeatother
\begin{document}

\preprint{AIP/123-QED}

\title[]{Triple Evaporation of Bialkali Antimonide Photocathodes and Photoemission Characterization at the PhoTEx Experiment}

% Force line breaks with \\
\author{J. Dube}
    \affiliation{Helmholtz-Zentrum Berlin für Materialien und Energie GmbH, Hahn-Meitner-Platz 1, 14109, Berlin, Germany}
    \affiliation{Humboldt-Universität zu Berlin, Institut für Physik, Newton-Straße 15, 12489, Berlin, Germany}
    \email{jonas.dube@helmholtz-berlin.de}
     
\author{J. Kühn}% 
    \affiliation{Helmholtz-Zentrum Berlin für Materialien und Energie GmbH, Hahn-Meitner-Platz 1, 14109, Berlin, Germany} 
    \email{julius.kuehn@helmholtz-berlin.de}

\author{C. Wang}
\affiliation{Helmholtz-Zentrum Berlin für Materialien und Energie GmbH, Hahn-Meitner-Platz 1, 14109, Berlin, Germany}
\affiliation{Universität Siegen, Institut für Werkstofftechnik, Paul-Bonatz-Straße 9-11, 57076, Siegen, Germany}

\author{S. Mistry}
\affiliation{Helmholtz-Zentrum Berlin für Materialien und Energie GmbH, Hahn-Meitner-Platz 1, 14109, Berlin, Germany}

\author{G. Klemz}
\affiliation{Helmholtz-Zentrum Berlin für Materialien und Energie GmbH, Hahn-Meitner-Platz 1, 14109, Berlin, Germany}

\author{A. Galdi}
\affiliation{Helmholtz-Zentrum Berlin für Materialien und Energie GmbH, Hahn-Meitner-Platz 1, 14109, Berlin, Germany}
\affiliation{Dipartimento di Ingegneria Industriale, Università degli Studi di Salerno, 84084, Fisciano (SA), Italy}

\author{T. Kamps}
\affiliation{Helmholtz-Zentrum Berlin für Materialien und Energie GmbH, Hahn-Meitner-Platz 1, 14109, Berlin, Germany}
\affiliation{Humboldt-Universität zu Berlin, Institut für Physik, Newton-Straße 15, 12489, Berlin, Germany}

\date{\today}% It is always \today, today,
             %  but any date may be explicitly specified

\begin{abstract}
The development of high-performance photocathodes is essential for generating high-brightness electron beams required by existing and future accelerators. This work introduces a state-of-the-art triple evaporation growth system designed for bialkali antimonide photocathodes. By enabling the simultaneous deposition of all three materials, this system significantly enhances vacuum stability and the reproducibility of photocathode fabrication. Complementing this, the novel characterization system PhoTEx allows spatially and spectrally resolved measurements of key photocathode parameters, such as quantum efficiency (QE), mean transverse energy (MTE), reflectance and lifetime. Crucially, all measurements are performed within a single compact setup, without moving the sample, preserving ultra-high vacuum conditions. The spectral resolved measurement of the reflectance allows the investigation of the color. Photocathode colorimetry may provide valuable insights into material homogeneity and aging. A Na-K-Sb photocathode was grown using the triple evaporation method, achieving an initial QE of $5.5\,\%$ at $520\,$nm. The photocathode was characterized at PhoTEx over two months, demonstrating consistent MTE measurements and a dataset with spectral response, reflectance and colorimetry data. Together, the triple evaporation growth system and PhoTEx mark a significant advancement in optimizing photocathodes with exceptional performance, paving the way for brighter and more stable electron sources for next-generation accelerator facilities.
\end{abstract}

\maketitle

\section{Introduction}
\label{sec:Introduction}
The generation of high-brightness electron beams plays a significant role for advanced accelerator facilities such as energy recovery linacs (ERLs)\cite{zhao2021,neumann2022}, free electron lasers (FELs)\cite{FEL_SLAC, schaber2023} and ultra-fast electron diffraction (UED) systems\cite{filippetto2022,zewail2006,Esuain2022_UED}. Since the performance of all these systems is predominantly determined by the electron source, photocathodes are used to meet the high requirements. The electron yield and the intrinsic emittance of the photocathode define the brightness limits for the accelerator.\cite{rao2014} These properties are described by the quantum efficiency (QE) and the mean transverse energy (MTE). Together with the lifetime they are the physical key parameters to describe the performance of a photocathode. While the QE defines the electron yield, the MTE is closely related to the normalized emittance $\epsilon_{x,y}$ of the emitted electron beam as shown in Eq. \ref{eq:emittance}. Here $\sigma_{x,y}$ describes the spot size of the illuminating light at the photocathode surface, and $m_e$ and $c$ are the electron mass and the vacuum speed of light, respectively.\cite{Lee2015}
\begin{eqnarray}
    \label{eq:emittance}
    \epsilon_{x,y} = \sigma_{x,y}\sqrt{\frac{\text{MTE}}{m_ec^2}}
\end{eqnarray}
Improvements here will have substantial impact on the performance of the accelerator, for example a lower emittance leads to a shorter gain length at FEL applications.\cite{Schmueser2014} Another important parameter is the reflectance of the photocathode surface, as it influences the QE of the photocathode.\cite{saha2023} The reflectance should be low because a reflected photon cannot contribute to the photoemission process.

Bialkali antimonide photocathodes, such as Cs-K-Sb, Rb-K-Sb or Na-K-Sb offer a high QE, low MTE and can photoemit when illuminated with visible light.\cite{dowell2010} Hence, they are suitable candidates to produce high-brightness electron beams.

Sometimes the photocathode growth system is attached to the photoinjector, more often it is in a separate laboratory for more flexibility. In this case a vacuum transport system is required, due to their high sensitivity to residual gases.\cite{wang2017,wang2021} Therefore, the photocathodes must remain under ultra-high vacuum conditions with at least low $10^{-9}\,$mbar during growth, characterization and operation in a photoinjector, including the transfer between the different systems.\cite{schmeisser2018} \\

At Helmholtz-Zentrum Berlin (HZB), bialkali antimonide photocathodes are developed and investigated for several years for high average current and low emittance ERL applications.\cite{schmeisser2018, cocchi2019, mistry2022} The photocathodes were prepared using the traditional method of sequential deposition of each element or co-deposition of the alkali metals and optimized with respect to their QE and lifetime.\cite{mistry2023,rozhkov2024} In 2024 a newly designed preparation chamber for the growth of bialkali antimonide photocathodes by triple evaporation of the three elements was commissioned.

Simultaneously, a new instrument has been set up, to measure all physical key parameters of the photocathodes, including the MTE. The system is based on the method of acceleration and subsequent free expansion of the electron beam.\cite{Lee2015} The MTE can be determined from the size of the electron beam at the detector. There are existing instruments based on this principle such as the TEmeter from Lee \textit{et al.}\cite{Lee2015} and the momentatron presented by Feng \textit{et al.}\cite{Feng2015}. A modified version is the TESS system developed by Jones \textit{et al.}\cite{jones2022} and a new system is under construction by Sertore \textit{et al.}\cite{sertore2022}. Based on these systems, we developed a novel instrument which is presented here. Additionally, this instrument is able to measure the QE and reflectance in one compact setup and can be used for lifetime studies. \\

The paper is structured in four parts: After an overview about the boundary conditions for the photocathode laboratory, the new triple evaporation growth system is described, followed by a description of the MTE measurement device - PhoTEx. In the last part, the growth of the first triple evaporation Na-K-Sb photocathode and its characterization at the PhoTEx instrument is presented.

\section{Photocathodes for an SRF gun}
The photocathodes grown at HZB are developed and investigated for use in the SRF photoinjector of the Superconducting RF Electron Accelerator Laboratory (SEALab, formerly bERLinPro).\cite{abo-bakr2020,neumann2022, kamps2022} Using an SRF photoinjector places high demands on the thermal and electric properties of the photocathode and the plug. A cylindrical plug with $10\,$mm diameter is used as substrate for the growth of the photocathodes. The design is specially adapted for the SRF photoinjector of SEALab and was extensively tested for operation.\cite{kuhn2019} It is made of molybdenum to meet the high thermal and electrical requirements and is optimized to avoid field emission in strong acceleration fields.\cite{kuhn2017, mistry2023} During operation, electric fields up to $20\,$MV/m may be applied to the cathode with an operating temperature between $80\,$K and room temperature.

\begin{figure}
    \centering
    \includegraphics[width=0.8\linewidth]{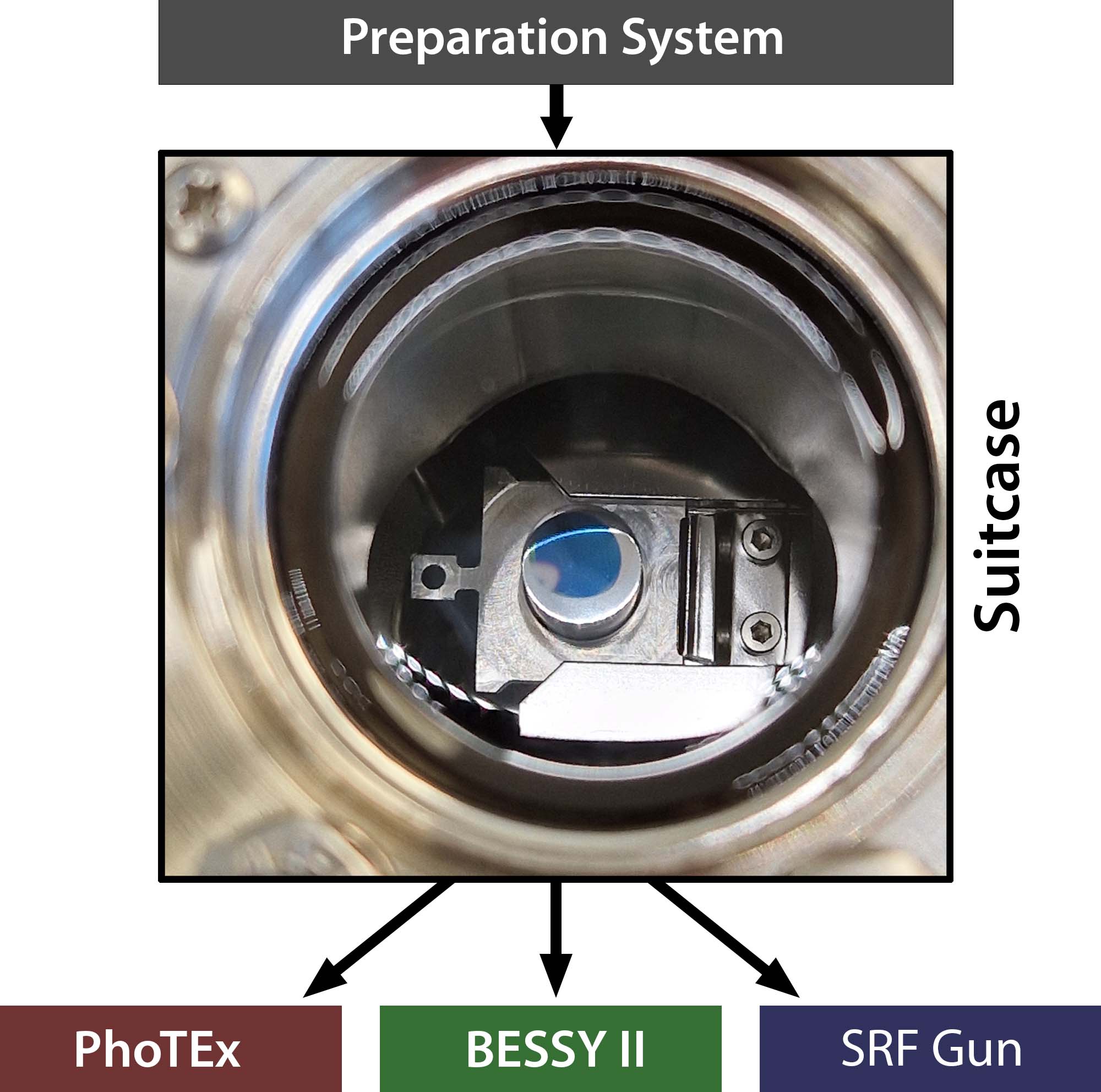}
    \caption{Omicron sample holder with the photocathode TWH13 in the suitcase; A suitcase is used to transfer samples from the photocathode preparation chamber to the SEALab SRF gun or characterization systems like PhoTEx and beamlines at the synchrotron radiation source BESSY II.}
    \label{fig:cathode_schema}
\end{figure}

The photocathodes are developed and tested on the same substrate that will later be used in the photoinjector. The plugs are mounted on an adapted Omicron flag style sample holder for preparation, characterization and transfer as shown in Fig. \ref{fig:cathode_schema}. At the photoinjector, the plug is removed from the sample holder and inserted into the injector.\cite{kuhn2017}
During the transfer, the photocathodes must remain particulate free and under ultra high vacuum conditions of at least low $10^{-9}\,$mbar due to their high chemical reactivity. A vacuum suitcase was built to transfer the samples between different facilities such as the preparation system, the SRF gun or characterization systems like PhoTEx or beamlines at the BESSY II synchrotron radiation facility at HZB.

The use of the photocathode in an SRF gun sets the boundary conditions for the transfer, characterization and the preparation of the photocathodes.

\section{Triple Evaporation Growth System}
\label{sec:growth_system}
A new photocathode preparation chamber was commissioned in summer 2024 to provide more flexibility in photocathode growth. In our previous setup, Sb and the alkali metals could not be deposited together. First, a Sb-layer was deposited and the alkali metals were evaporated sequentially or co-deposited afterwards.\cite{schmeisser2018,mistry2023} This limited the ability to develop new growth recipes. Sb was evaporated from an effusion cell, while SAES dispensers were used for the alkali metals. Therefore, parts of the vacuum chamber had to be regularly vented to change the dispensers. \\

\begin{figure}
    \centering
    \includegraphics[width=\linewidth]{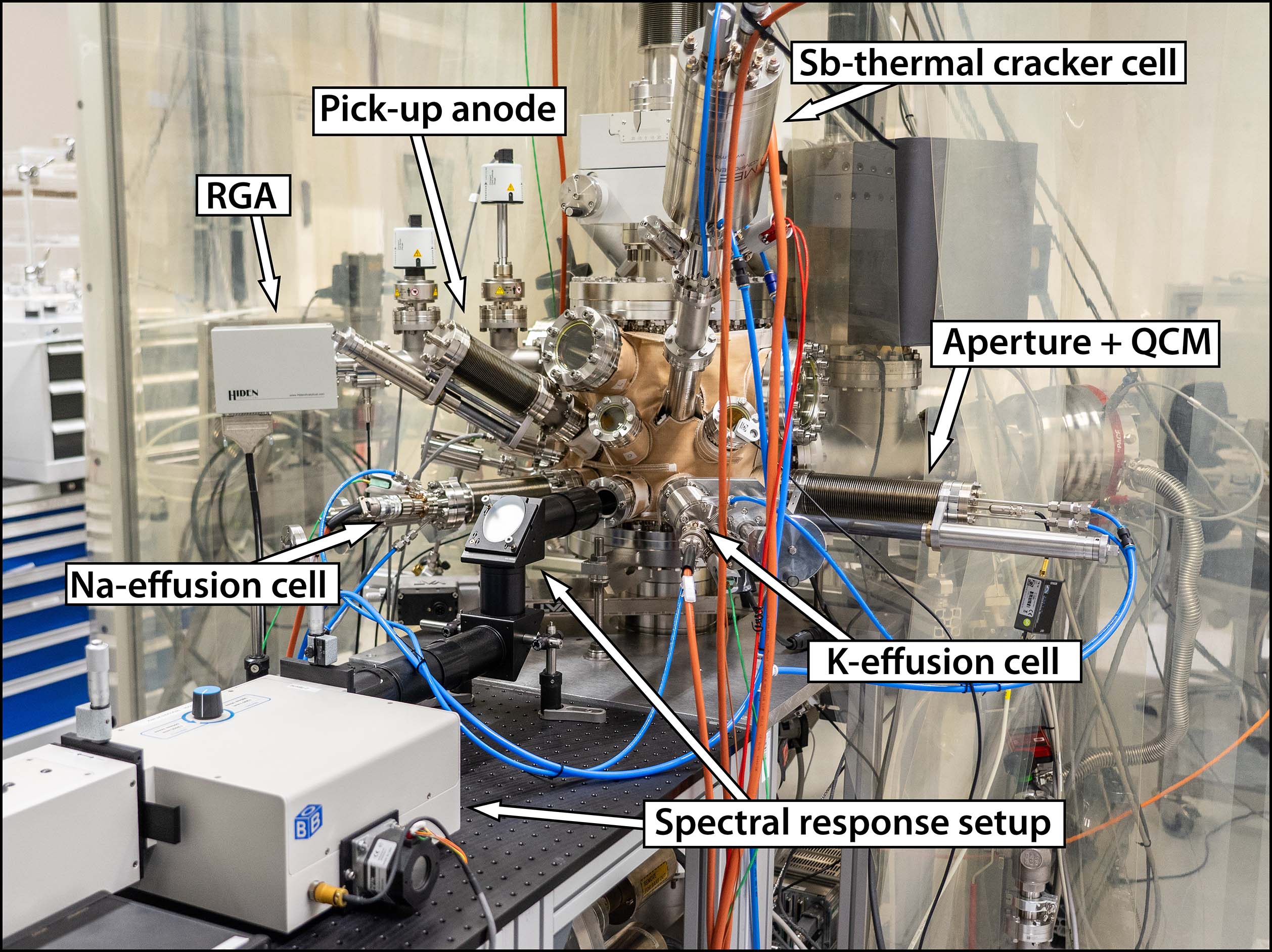}
    \caption{Photocathode laboratory at HZB. The new triple evaporation growth chamber is shown with the spectral response setup at the front.}
    \label{fig:prep_chamber}
\end{figure}

In the new setup the alkali dispensers were replaced by effusion cells filled with SAES alkali metal pills. The antimony is evaporated from beads in a thermal cracker cell. The three sources are oriented so that they all point towards the substrate position to allow simultaneous evaporation of the three materials. The thermal cracker cell for antimony is placed above the photocathode at an angle of $52^\circ$ at a fixed distance of $148\,$mm as shown in Fig. \ref{fig:prep_chamber}. The alkali effusion cells are located in the horizontal plane at an angle of $30^\circ$ to the normal of the photocathode surface. The distance of the alkali metal sources can be adjusted with a Z-drive and is typically set to $60\,$mm. The photocathode is placed on a heatable XYZ$\Phi$ manipulator. A circular $8\,$mm aperture is placed in front of the plug to achieve a well defined photocathode area on the substrate. The aperture has a bevelled edge to reduce shadowing effects from the sources and increase the area of full overlap of all three materials. The deposition rate can be measured using a water cooled quartz crystal microbalance (QCM) placed in the same position as the plug during the calibration measurement. The rate can be adjusted by the temperature settings of the sources. Additionally, an argon sputter gun is attached to the system for the cleaning of the plug surface.

The photocathode can be illuminated directly from the front with the spectral response setup. A xenon lamp and a monochromator are used to generate monochromatic light with a bandwidth of $5\,$nm in the full visible regime. A pick-up anode is placed close to the photocathode and is connected to a picoammeter to measure the extracted current. The calculation of the QE is described in more detail in Sec. \ref{sec:spectral_response}. During the growth of a photocathode, the light source is changed to a $520\,$nm laser with a maximum output of $0.9\,$mW, to increase the sensitivity especially at the beginning of the growth. The light power can be adjusted with neutral density filters.

The vacuum is maintained by a turbo pump, an ion getter pump (IGP) and a non-evaporable getter (NEG) pump. The partial pressure of any residual gases can be measured with a residual gas analyzer (RGA). After bake out, a base pressure of $9\times10^{-11}\,$mbar was achieved. The preparation system is attached to an analysis chamber for X-ray photoelectron spectroscopy (XPS) measurements\cite{schmeisser2018} and to a load lock to insert new samples. A second load lock is located at the back of the vacuum system to transfer the photocathodes to other facilities such as PhoTEx or the photoinjector of SEALab.

\subsection{Suitcase for Sample Transfer}
\label{sec:suitcase}
\begin{figure}
    \centering
    \includegraphics[width=\linewidth]{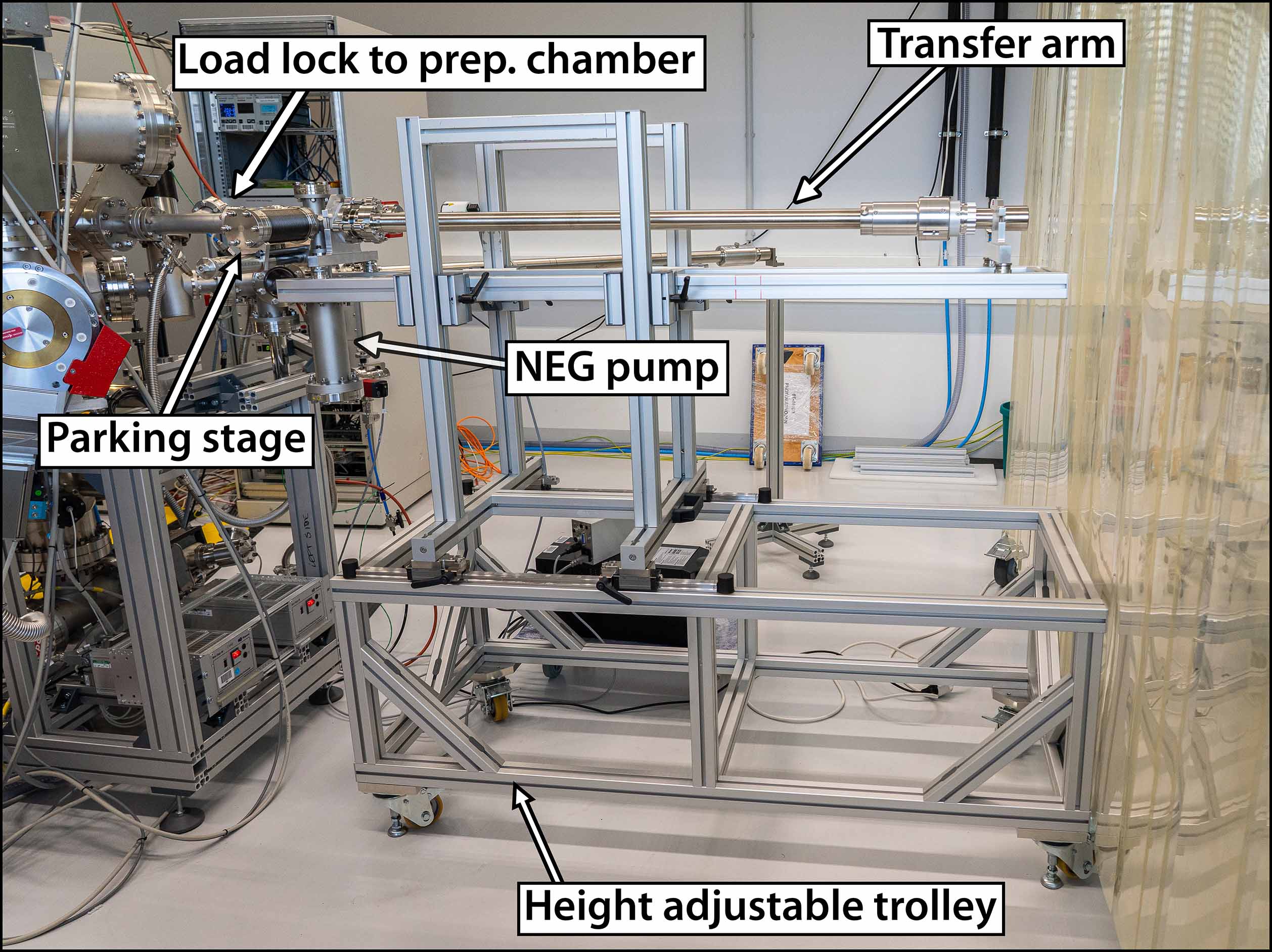}
    \caption{The vacuum suitcase attached to the preparation chamber in the photocathode laboratory.}
    \label{fig:suitcase}
\end{figure}

A new suitcase was built, to transfer the samples at ultra high vacuum conditions and is shown in Fig. \ref{fig:suitcase}. The suitcase is equipped with a pincer connected to a transfer arm. Therefore, no transfer mechanism is required at the receiving laboratory. Hence, it is compatible with most vacuum systems. A parking stage allows the simultaneous transfer of up to three samples. A non evaporable getter (NEG) pump is used to achieve a base pressure of $3\times 10^{-11}\,$mbar. Only during the gripping of a sample the pressure may temporarily increase due to degassing of the transfer arm. The pressure can be maintained below $3\times10^{-9}$ by moving the transfer arm very slowly. An additional IGP can be used to increase the pumping speed. During a transfer to another laboratory, the IGP and the vacuum gauge are connected to an uninterruptible power supply. The whole suitcase is mounted on wheels and is height adjustable. Therefore, it can be flexibly adjusted to different systems.

\section{Characterization System PhoTEx}

\label{sec:photex}
The \textbf{Pho}toemission and \textbf{T}ransverse \textbf{E}nergy E\textbf{x}periment - PhoTEx is an instrument to characterize the physical key parameters of photocathodes for electron accelerators. Besides the MTE, also the QE, reflectance and lifetime can be investigated at PhoTEx. All parameters can be measured spectrally resolved. This is of major importance as the photoemission process is primarily determined by the energy of the incident photons $\hbar \omega$ and the effective workfunction of the material $\phi_\text{eff}$.\cite{spicer1958,dowell2009,saha2023} The difference of both values is defined as the excess energy $E_\text{excess} = \hbar \omega - \phi_\text{eff}$. As a first approximation, the MTE and QE can be described by the excess energy as shown in Eq. \ref{eq:QE_MTE_theory}.\cite{saha2023} The equation for the QE is valid close to the threshold energy, while the MTE formula is an approximation for high excess energies.
\begin{eqnarray}
\label{eq:QE_MTE_theory}
    \text{QE} \propto E^2_\text{excess} \qquad \text{MTE} = \frac{E_\text{excess}}{3}
\end{eqnarray}
Deviations to this first approach may occur for semiconducting materials and at non-zero temperature.\cite{saha2023} Close to the threshold energy the MTE is no longer linear but saturates at the thermal limit $k_BT$, where $k_B$ is the Bolzmann constant.

\subsection{Experimental Setup}
\begin{figure}
    \centering
    \includegraphics[width=\linewidth]{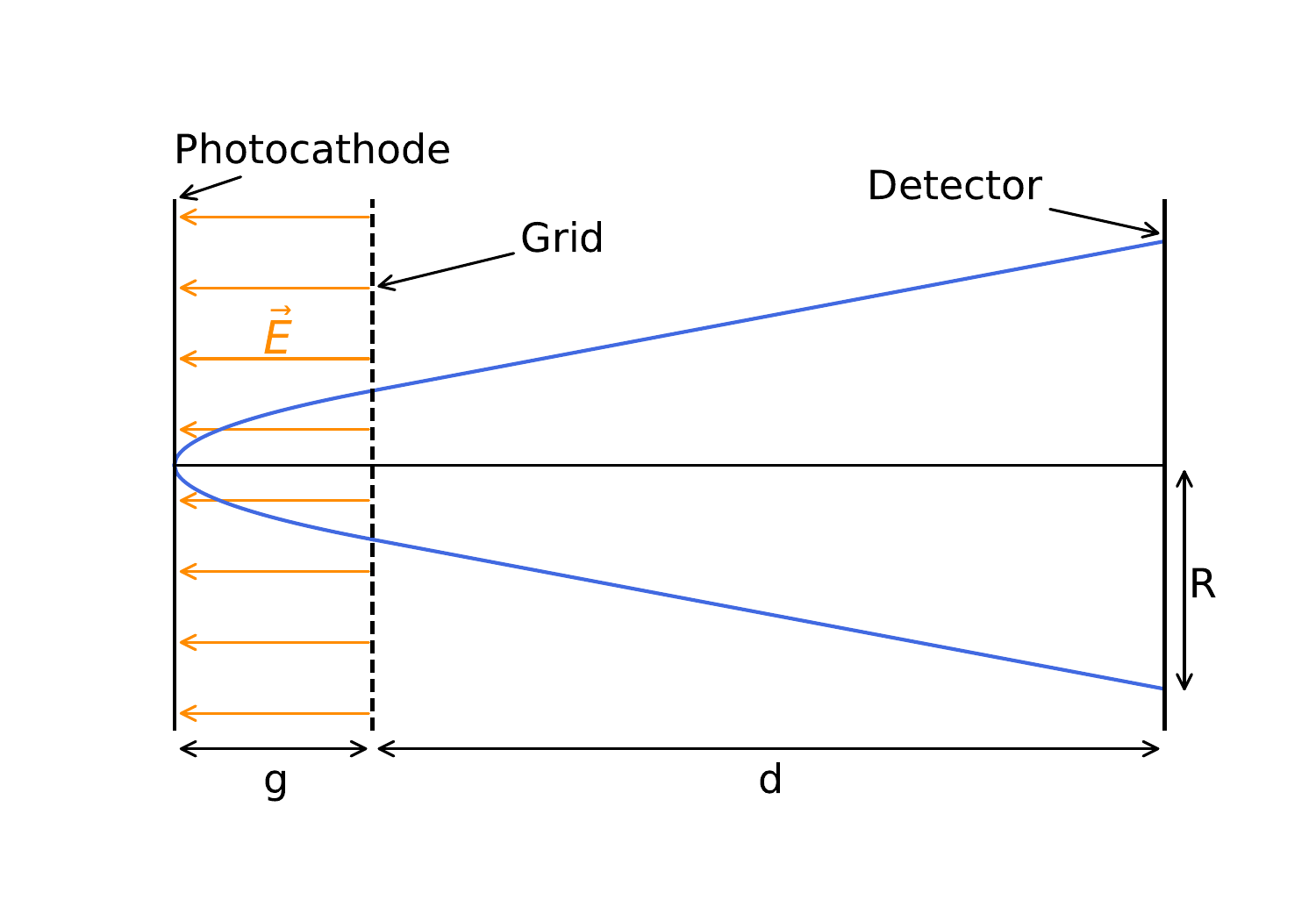}
    \caption{Schematic of the measurement principle of PhoTEx. The electrons are accelerated in a homogeneous E-field (orange) between the negatively biased photocathode and a grounded grid. After passing the grid, the electron beam expands freely until it hits the detector. The mean transverse energy can be determined from the mean offset at the detector $R$.}
    \label{fig:MTE_schema}
\end{figure}

For the MTE measurement, a negative voltage $U$ is applied to the photocathode and a conductive copper grid is placed at distance $g$ to achieve a homogeneous electric field as shown in Fig. \ref{fig:MTE_schema}. The grid has square holes with a size of $11\,\upmu$m and an open area of $55\,\%$. After the electrons pass through the grid they enter a field free region. The electron beam is expanding in this drift tube of length $d$ to increase the spatial resolution at the detector. With the assumption of a perfect homogeneous field and field free drift region, Eq. \ref{eq:MTE} can be derived to determine the MTE from the radius $R$ of the electron beam at the detector.\cite{sertore2022} \\
\begin{eqnarray}
\label{eq:MTE}
    \text{MTE}=\frac{R^2}{(2g+d)^2}Ue
\end{eqnarray}
The detector is a one stage microchannel plate (MCP) with a gain up to $10^4$ and a P43 phosphor screen. A characterization of the detector in our laboratory showed a doubling of the intensity every $36\,$V.

\begin{figure}
    \centering
    \includegraphics[width=\linewidth]{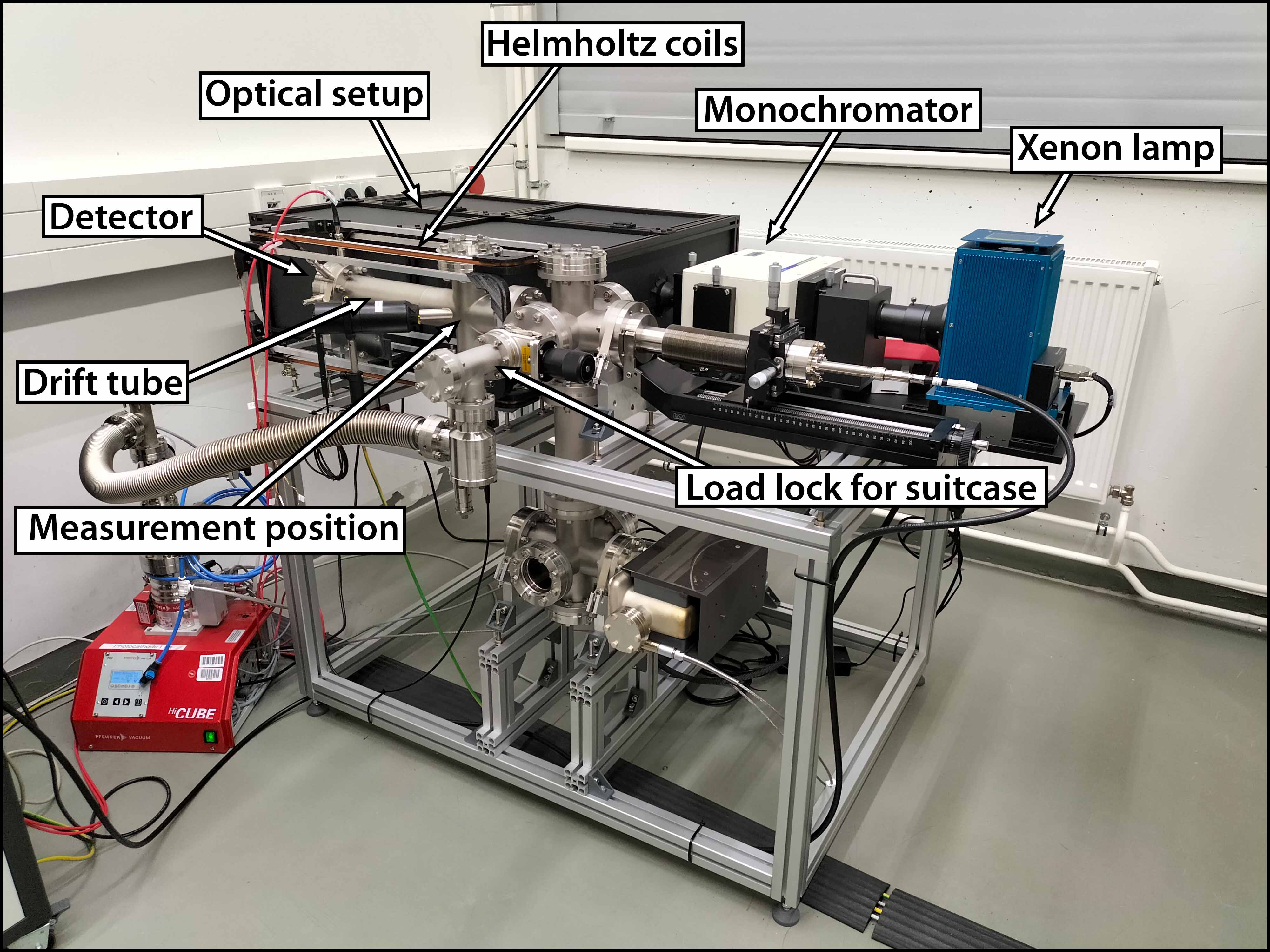}
    \caption{PhoTEx assembled in the MTE laboratory at HZB}
    \label{fig:PhoTEx_photo}
\end{figure}

The whole PhoTEx setup is shown in Fig. \ref{fig:PhoTEx_photo} and contains three different parts: the vacuum chamber, an optical setup and Helmholtz coils to steer the electron beam on the detector. A photocathode can be transferred to PhoTEx with the vacuum suitcase mentioned above. The suitcase is attached to the load lock of PhoTEx which can be separately pumped with a pumping cart. The photocathode is moved with an XYZ manipulator to the measurement position at $10\,$mm distance in front of the grid. The XY-stage of the manipulator allows spatially resolved characterization of the photocathode. A non evaporable getter pump and an ion getter pump ensure a vacuum base pressure of $5\times 10^{-11}\,$mbar. A copper rod is mounted on a Z-drive below the measurement position. When inserted, it is placed between the photocathode and the grid and acts as a pick-up anode to measure the quantum efficiency.

\begin{figure*}
    \centering
    \includegraphics[width=\linewidth]{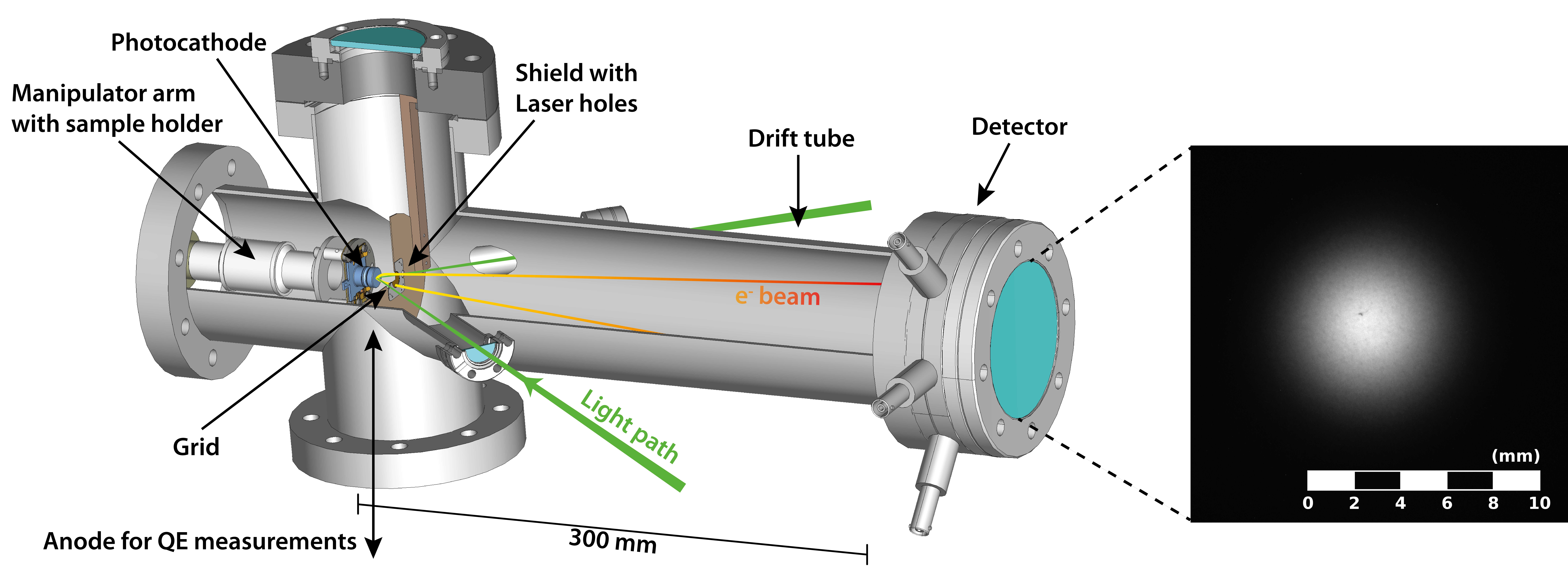}
    \caption{Illustration of the analysis chamber of PhoTEx. The 3D model is sectioned for a better visualization. A real image of an electron beam at the detector is shown on the right. An anode can be moved in from the bottom with a Z-drive for QE measurements.}
    \label{fig:cut_PhoTEx}
\end{figure*}

The grid with $8\,$mm diameter is surrounded by a molybdenum (Mo) shield with $64\,$mm diameter. This enables a better shielding for the drift tube and a more homogeneous electric field in the acceleration gap. The incident and reflected light passes through holes in the Mo-shield with $3.5\,$mm diameter as shown in Fig. \ref{fig:cut_PhoTEx}. With that we achieve an illumination angle of $38.7^\circ$. However, the optical path is very sensitive to misalignments. \\

\begin{figure}
    \centering
    \includegraphics[width=0.9\linewidth,trim= 20 75 120 70,clip]{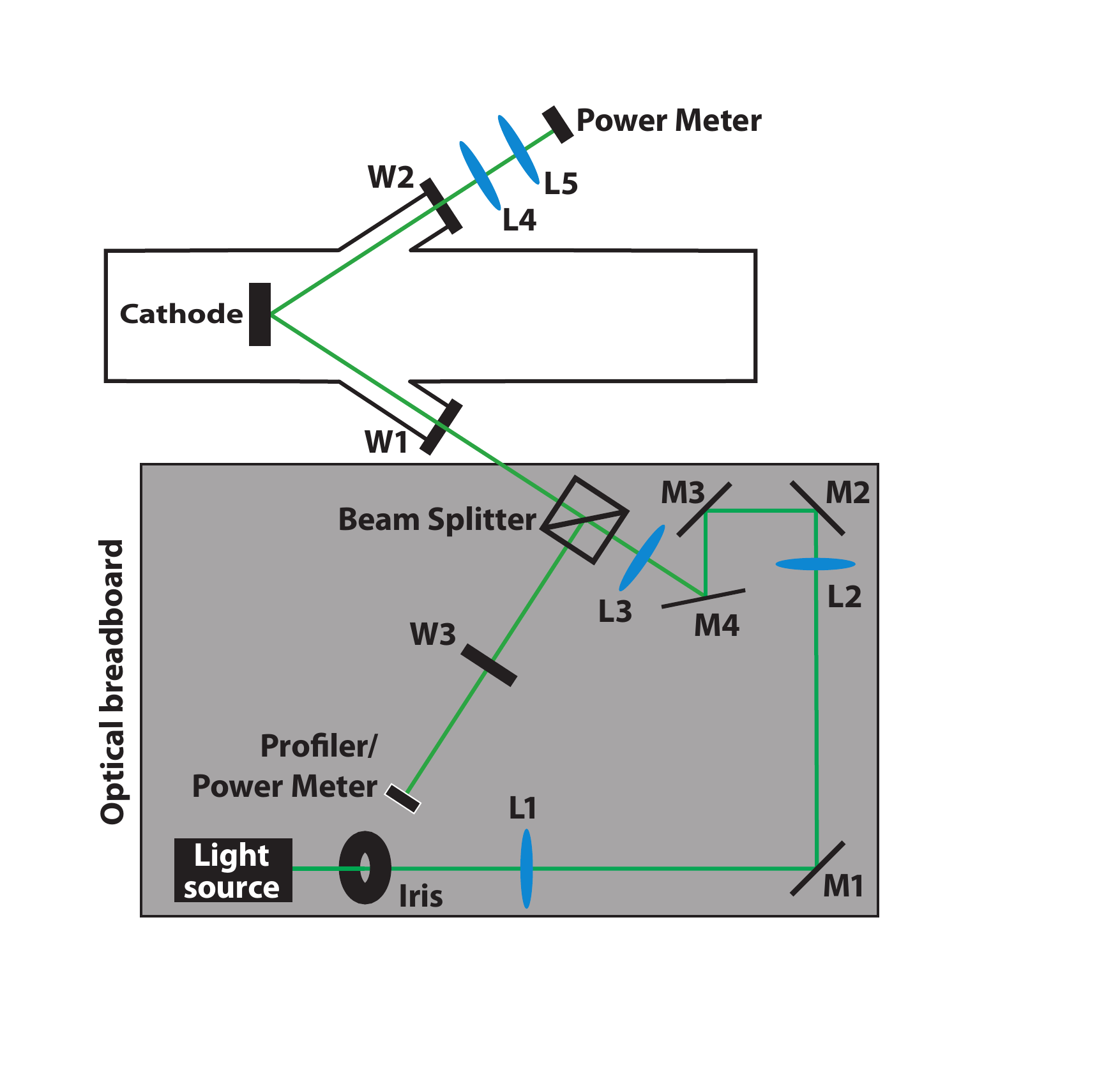}
    \caption{Schematic of the optical setup at PhoTEx. Three Lenses (L) are used to image an iris onto the photocathode. Mirrors (M) are used to fold the optical path. The light enters and exits the vacuum chamber through vacuum windows (W). An identical window is placed in the reference light path (W3).}
    \label{fig:optical_setup}
\end{figure}
The light source used at PhoTEx is a $75\,$W Xenon lamp connected to a monochromator. This allows the illumination of the photocathode with monochromatic light in the regime of $(400-700)\,$nm with a bandwidth of $5\,$nm. Three lenses are used to image an iris aperture onto the photocathode surface with a demagnification of 10:1. A beam splitter is placed just before the light enters the vacuum chamber to allow live monitoring of the optical power as shown in Fig. \ref{fig:optical_setup}. An identical vacuum window is placed in this reference light path to achieve the same conditions as at the photocathode. The spot size of the light beam can be measured using a beam profiler which is located at the reference light path. The lens system is adjusted so that the photocathode and the beam profiler are located at the focal point with a deviation of $\pm\,2\,$mm. The $1\sigma$ radius of the Gaussian light spot was measured in this area and is $\sigma = (43\pm3)\,\upmu$m. Since the photocathode is placed at the focal point, the reflected beam is strongly divergent. Two additional lenses are used to focus the reflected beam onto the second power meter to measure the reflectance. The optical density of these lenses were measured and is considered in the calculation of the reflectance. Neutral density (ND) filters can be placed in the optical path to reduce the optical power. A reduced optical power leads to a lower photocurrent and therefore reduces space charge effects. In the best case the number of electrons in the system at the same time $N_e$ is close to $1$ or even smaller to avoid electron-electron interaction. This can be estimated with Eq. \ref{eq:number_electrons} from the flight time $\tau$ which is given by the system parameters $g$, $d$, $U$ and the electron mass $m_e$ and charge $e$.
\begin{eqnarray}
\label{eq:number_electrons}
\begin{split}
N_e &= \frac{I}{e}\tau = \text{QE} \frac{P_\gamma \lambda}{hc} \tau \\
\tau &= -\frac{gm_ev_{z0}}{Ue}+\sqrt{\left( \frac{gm_ev_{z0}}{Ue}\right)^2 + \frac{2g^2m_e}{Ue}} + \frac{d}{\sqrt{v_{z0}^2+\frac{2Ue}{m_e}}} \\
\end{split}
\end{eqnarray}
This equation is derived with the non-relativistic approach, which is a very good approximation for the acceleration voltages used at PhoTEx. The initial longitudinal velocity of the electron $v_{z0}$ is small compared to the acceleration and can be neglected for rough approximation. The photocurrent $I$ is derived from a QE measurement as described in Eq. \ref{eq:QE}. \\

The electrons are very sensitive to external fields due to their low kinetic energy. The magnetic field strength of the Earth would already be enough that the electrons are not hitting the detector anymore.
Helmholtz coils are used to compensate external magnetic fields at PhoTEx. Only the magnetic field components perpendicular to the velocity are deflecting the electrons. Simulations confirmed, that the longitudinal component of the magnetic field is negligible. Therefore, only the transverse field components have to be compensated. Two pairs of rectangular Helmholtz coils are placed around the analysis chamber at PhoTEx. The coils create a homogeneous magnetic field in the acceleration gap and the drift tube and allow steering of the electron beam to the center of the detector. The technical data of the full PhoTEx setup is summarized in Tab. \ref{tab:technical_data}.

\begin{table}
\caption{Technical data of PhoTEx. Listed are the parameters for the
vacuum system, optics and magnetic shielding.}
\label{tab:technical_data}
\begin{ruledtabular}
\begin{tabular}{r l @{\hspace{0em}}}
Sample holder: & Omicron flag style \\
Base pressure: & $5\times 10^{-11}$\,mbar \\
Acceleration gap: & 10\,mm \\
Drift distance: & 300.9\,mm \\
Acceleration voltage: & < 6.5\,kV \\
Max. E-field: & 0.65\,MV/m \\
Grid: & 11\,$\upmu$m hole size, 55\,\% open area \\
Detector: & One stage MCP + P43 phosphor screen \\
\hline
Light source: &  Zolix tunable light source TLS2-X75A-G; \\
 & 75W Xenon lamp + Monochromator \\
 Spectral regime: & (400-700)\,nm \\
 Bandwidth: & 5\,nm \\
 Spot size at cathode: & 1$\sigma$ diameter $=(86\pm6)\,\upmu$m  \\
 Illumination angle: & 38.7$^\circ$ \\
 \hline
Magnetic shielding: & Helmholtz coils for transvserse directions
\end{tabular}
\end{ruledtabular}
\end{table}

\subsection{Particle Tracking Simulation}
\label{sec:IF}
The assumption of a perfect homogeneous electric field between the cathode and the grid and a perfect field free drift region cannot be achieved in reality. The measurement of the MTE was simulated to quantify the impact of field imperfections and determine an instrument function to correct for these deviations.

\begin{figure}
    \centering
    \includegraphics[width=0.8\linewidth]{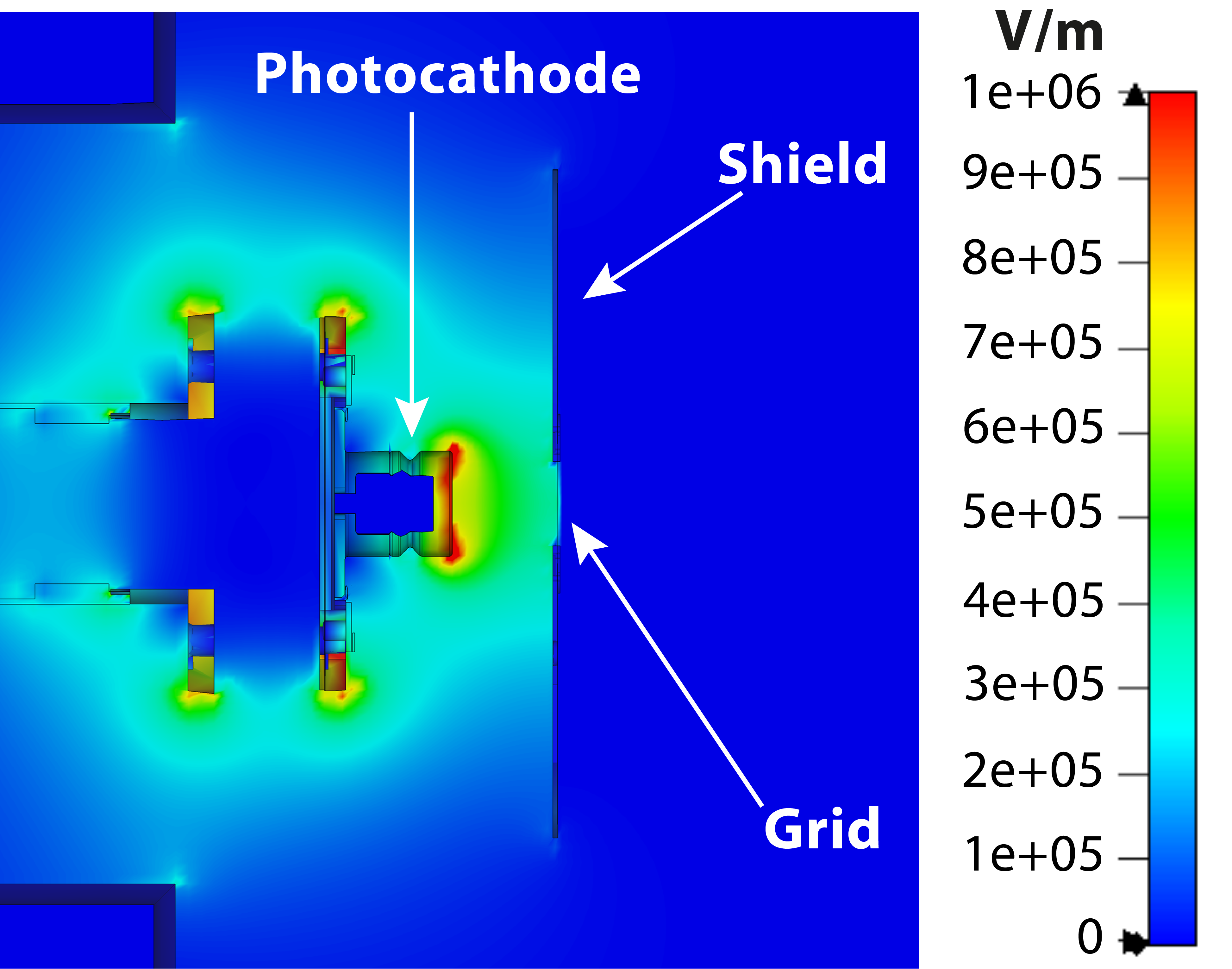}
    \caption{Simulation of the electric fields in the analysis chamber of PhoTEx. The field between the photocathode and the grid is not perfectly homogeneous. A good electrical shielding to the drift tube is achieved by the grid and the shield.}
    \label{fig:E-field}
\end{figure}

The electric field in the measurement region was simulated with CST Studio Suite\cite{CST} on a three dimensional grid. A constant voltage of $5000\,$V was applied to the surface of the photocathode. The grid, vacuum chamber and detector have a potential of $0\,$V. In case of a perfect homogeneous field we would expect a constant field strength of $0.5\,$MV/m between the photocathode and the grid. As shown in Fig. \ref{fig:E-field} the field is decreasing from $0.69\,$MV/m at the photocathode surface to $0.38\,$MV/m at the grid in reality. This will also influence the trajectories of the emitted electrons and hence the spot size of the electron beam at the detector. 

Electron distributions with a specified MTE were tracked through the simulated E-fields with the particle tracking algorithm ASTRA\cite{flottmann2003} to simulate the electron distribution at the detector. This simulation was performed for a range of different start values i.e. different voltages, MTE's and light spot sizes at the photocathode surface. Each time a distribution of 100\,000 electrons was tracked through the simulated electric fields inside PhoTEx. A radial intensity function of the generated detector image was calculated to determine the beam radius, as described in the next section. This simulated measurement $R_\text{meas}$ was then compared with the theoretically expected value $R$ from Eq. \ref{eq:MTE}. The relation between the measured value and the expected value is generally linear, with a gradient of $\alpha \approx 0.87$, meaning that $R$ is about $13\,\%$ smaller than the measured value. This is due to the imperfection in the electric fields shown in Fig. \ref{fig:E-field}, which lead to a defocusing of the electron beam.

The theory in Eq. \ref{eq:MTE} assumes a point-like emission area of the electrons, but in reality the electron distribution at the detector is a superposition of the transverse energy and the initial spatial distribution of the electron i.e. the light spot size at the photocathode surface. Hence, the measured radius is saturating for small radii. This effect is negligible if the spot size of the electron beam at the detector is much larger then the light spot at the photocathode surface. Therefore, the light spot at the cathode should be small. At PhoTEx the $1\sigma$ radius of the spot at the photocathode surface is $\sigma = (43\pm3)\,\upmu$m. The simulations showed that this spot size is sufficiently small that the relation between the measured radius and the expected value can be assumed to be linear. The instrument function in Eq. \ref{eq:Instrument_function} was developed to correct the measured radius for this specific spot size of the illuminating light. The parameters $\alpha$ and $\beta$ were found by fitting this function to the simulated data.
\begin{eqnarray}
    \label{eq:Instrument_function}
    \begin{split}
    R &= \alpha \times R_\text{meas} - \beta \\ 
    \alpha &= 0.873 \qquad \beta = 1.61\times 10^{-4}\,\text{m}
    \end{split}
\end{eqnarray}
The deviation between the measured and the real value can be significantly reduced by correcting the measurement with the instrument function. The remaining uncertainty $U_R$ is dependent on the measured radius and is listed below.
\begin{equation*}
    \begin{split}
        R_\text{meas} < 2.5\,\text{mm} \qquad &\rightarrow \qquad U_R = 4\,\% \\
        2.5\,\text{mm} < R_\text{meas} < 11\,\text{mm} \qquad &\rightarrow \qquad U_R = 2\,\% \\
        11\,\text{mm} < R_\text{meas} \qquad &\rightarrow \qquad U_R = 4\,\% \\
    \end{split}
\end{equation*}

\subsection{Data Acquisition and Analysis}
\subsubsection{Spectral Response}
\label{sec:spectral_response}
During a spectral response measurement the iris aperture can be opened to increase the optical power and therefore improve the signal-to-noise ratio. A wavelength scan is performed with a step-width of $5\,$nm and the optical power $P_\gamma$ and the photocurrent $I_{e}$ are measured simultaneously. The photocurrent is averaged over 20 data points to reduce noise. The standard deviation of the measurements is the largest contribution to the total uncertainty of the measured QE, especially for low currents. The optical background power and the darkcurrent are measured and subtracted from the measurement. The QE is then calculated for each wavelength $\lambda$ with Eq. \ref{eq:QE}. \cite{rao2014} Here $N_{e}$ is the number of electrons extracted, $N_\gamma$ the number of incident photons, and $h$ and $c$ the Planck constant and the vacuum speed of light, respectively.

\begin{eqnarray}
\label{eq:QE}
    \text{QE} = \frac{N_{e}}{N_\gamma} = \frac{I_{e}hc}{P_\gamma \lambda e}
\end{eqnarray}

\subsubsection{Reflectance and Colorimetry}
For a reflectance measurement, the wavelength is scanned with the monochromator and the initial and reflected optical power is measured with two power meters. The reflectance is calculated from the ratio of both measurements. Here, also the transmission of the viewport and lenses used to focus the reflected beam onto the second power meter are taken into account. 

From the measured reflectance curve, the color of the photocathode can be derived using the CIE 1931 color space. It is merely a distinct mode of presentation and does not possess more information. However, it is a good way to make the reflectance data more tangible, as it corresponds to our human perception. With this method the color of the photocathodes can be studied in a systematic way. It may provide valuable insights into the homogeneity of the photocathode film and could facilitate a preliminary assessment of the photocathode quality in the future.

The tristimulus values $X$, $Y$ and $Z$ are calculated from the measured spectral reflectance $R_\lambda$ using the CIE color matching functions $\bar{x}$, $\bar{y}$ and $\bar{z}$ and the spectral power distribution $S$ of a specified illuminant using Eq. \ref{eq:color}.\cite{Kruschwitz2018} The sum is performed from $(400-700)\,$nm in steps of $5\,$nm. The standard illuminant D65 and the CIE 1931 color-matching functions for a 2 degree observer are used for our data analysis.\cite{CIE1931_D65,CIE1931_cmf} $k$ is a normalization constant such that 1 is the maximum value for $Y$.
\begin{eqnarray}
\label{eq:color}
\begin{split}
X &= k \sum_\lambda S_\lambda R_\lambda \bar{x}_\lambda \\
Y &= k \sum_\lambda S_\lambda R_\lambda \bar{y}_\lambda \\
Z &= k \sum_\lambda S_\lambda R_\lambda \bar{z}_\lambda \\
k &= \frac{1}{\sum_\lambda S_\lambda \bar{y}_\lambda} \\
\end{split}
\end{eqnarray}
The tristimulus values can be converted to the sRGB color space for display on computer screens.

\subsubsection{Mean Transverse Energy}
\label{sec:data_analysis_MTE}

The MTE can be determined with Eq. \ref{eq:MTE} with $R$ the root-mean-square radius of the electrons. For the measurement, a data image is taken with the detector camera and corrected by subtraction of a dark image. The dark image is taken without illuminating the photocathode but with the same exposure time and applied voltages. A mask is applied to remove all pixels outside the detector area. 
A 3D Gaussian fit is applied to the resulting image with an independent variance in both directions. The radius of the electron distribution is then given by $R=\sqrt{\sigma_x^2+\sigma_y^2}$. This formula reduces to $R=\sqrt{2}\sigma$ with $\sigma= \sigma_x=\sigma_y$ as we expect a radial symmetry. From the 3D Gaussian the center of the distribution is known. A radial intensity function (RIF) is calculated by averaging the intensity of all pixels with equal distance to the center to reduce noise. A Gaussian fit is applied to the RIF and the Radius is determined as described in Eq. \ref{eq:Radius_sigma}. The conversion factor $f_c$ is used to convert the number of pixels into a distance.
\begin{eqnarray}
\label{eq:Radius_sigma}
    R = \sqrt{2}\sigma\times f_c
\end{eqnarray}
\begin{figure}
    \centering
    \includegraphics[width=\linewidth]{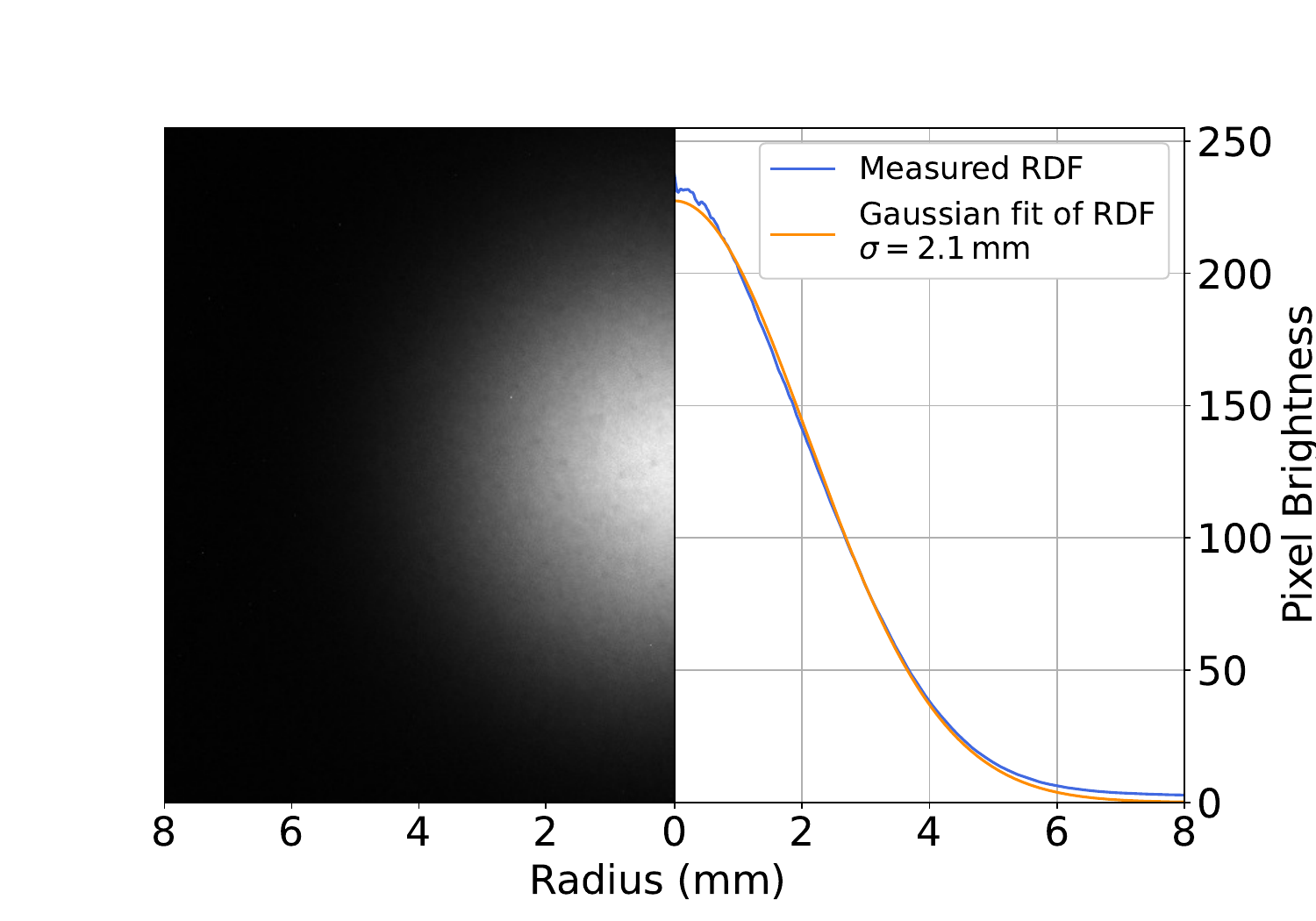}
    \caption{The radial intensity function (RIF) on the right is calculated from an image of the detector camera shown on the left. A conversion factor is applied to convert the number of pixels to a distance.}
    \label{fig:RIF}
\end{figure}
A detector image and the associated RIF is shown in Fig. \ref{fig:RIF}. The resulting radius is corrected by the instrument function Eq. \ref{eq:Instrument_function} derived in Sec. \ref{sec:IF}. The total uncertainty of the radius is given by the uncertainty of the instrument function discussed above, the resolution of the detector, the fit uncertainty of the RIF and the Davisson-Calbick lens effect. The latter describes the deflection of an electron at the edge of a hole at the copper grid due to the bending of electric fields. This angle deviation can be described with Eq. \ref{eq:Davisson_calbick}.\cite{Davisson1932, Yu2021} Here $D$ is the width of a grid cell and $g$ is the distance between the photocathode and the grid.
\begin{eqnarray}
\label{eq:Davisson_calbick}
    \Theta = \frac{D}{8g}
\end{eqnarray}
The deviation at the detector is then given by $\Delta R_2 = \tan(\Theta)d$, with $d$ the distance between the grid and the detector. As some of the errors are not randomly distributed but primarily lead to an increase of the measured radius, the total uncertainty of the radius $\Delta R_\text{tot}$ is the linear sum of all four sources of error.
\begin{table}
\caption{Overview of all considered error sources and their related uncertainty.}
\label{tab:total_uncertainty}
\begin{ruledtabular}
\begin{tabular}{r l @{\hspace{5em}}}
Detector resolution: & $\Delta R_1 = 25\,\upmu$m \\
Davisson-Calbick effect: & $\Delta R_2 = 41\,\upmu$m \\
Uncertainty RIF: & $\Delta R_3$ \\
Uncertainty instrument function: & $\Delta R_4 = U_R$ \\
\hline
Uncertainties distances: & $\Delta$d = 0.3\,mm \\
 & $\Delta$g = 0.1\,mm \\
Uncertainty voltage: & $\Delta$U = 0.1\,\% of measured value
\end{tabular}
\end{ruledtabular}
\end{table}

The MTE can be determined in two different ways. It can be calculated as a single shot measurement by directly inserting the measured radius in Eq. \ref{eq:MTE}. Nevertheless, the single shot measurement is subject to a high degree of uncertainty. A more profound measurement can be done with a voltage scan. In this case the radius at the detector is measured as a function of the applied voltage at the photocathode. The MTE is determined by fitting the data with Eq. \ref{eq:voltage_scan}. The MTE is the only fit parameter, since the voltage $U$, distances $g$ and $d$ and the electron charge $e$ are known.
\begin{eqnarray}
    \label{eq:voltage_scan}
    R = (2g+d)\sqrt{\frac{\text{MTE}}{Ue}}
\end{eqnarray}
The uncertainty of the radius is taken into account in the fit and is reflected in the uncertainty of the fit parameter. The uncertainties due to the voltage $\Delta U$ and the distances $\Delta g$ and $\Delta d$ are considered separately. The contribution of these values to the uncertainty of the MTE is calculated for each single measurement of the voltage scan. The maximum value is added linearly to the uncertainty of the fit parameter to give the total uncertainty $\Delta$MTE.

\section{Triple evaporation Na-K-Sb photocathode}
\subsection{Growth Procedure}
\label{sec:growth_procedure}
After the commissioning of the new growth chamber, a Na-K-Sb photocathode was grown with the triple evaporation approach for the first time at October 15, 2024. This photocathode will be referred to as TWH013 in the following. A full molybdenum plug was inserted to the preparation system, degassed and cleaned with argon sputtering. Before the growth, the deposition rate of all three sources were measured with the QCM to be $0.011\,$\r{A}/s for Sb, $0.029\,$\r{A}/s for Na and $0.19\,$\r{A}/s for K.

\begin{figure}
    \centering
    \includegraphics[width=\linewidth]{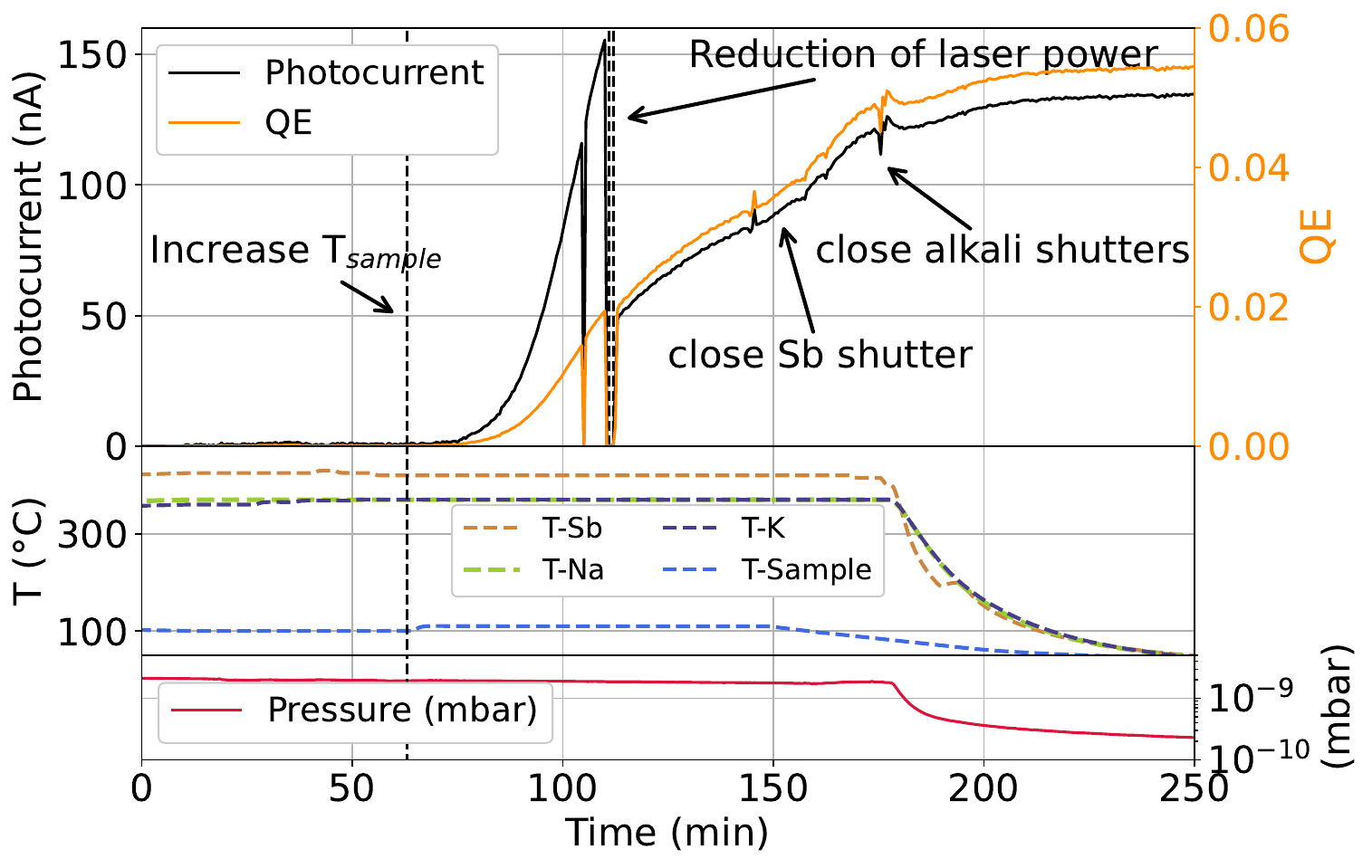}
    \caption{Growth procedure of the photocathode TWH013. An increase of the photocurrent started shortly after the sample stage temperature was increased to $110^\circ$C. Besides the photocurrent/QE, also the temperatures and the vacuum pressure is shown. The photocurrent was measured at $520\,$nm.}
    \label{fig:growth}
\end{figure}
In Fig. \ref{fig:growth} we show the growth procedure. The sample stage was heated to $100^\circ$C and the shutters of the Sb, K and Na sources were simultaneously opened at $t=0\,$min. No increase of the photocurrent could be observed within the first $60\,$min. Therefore, the stage temperature was increased to $110^\circ$C after $65\,$min, to enhance the chemical reaction of the materials. Shortly after, the photocurrent started to increase. The vacuum pressure during the deposition was stable at $2\times10^{-9}\,$mbar. The laser power was reduced with an additional ND $0.5$ filter after $112\,$min to lower the extracted charge during the growth. The dips after $105\,$min and $111\,$min are caused by temporarily switching of the power supply for the photocurrent measurement to check the darkcurrent and the photo response. At the end of the deposition, after $145\,$min, the shutters were opened and closed to maximize the photocurrent. Closing the alkali shutters reduced the photocurrent indicating an alkali deficiency, so they were opened again. The sample temperature was reduced and the Sb source was closed $12\,$min later. After $175\,$min the alkali shutters were closed and the heaters of the sources were turned off. While cooling down, the QE further increased and stabilized at QE$\,=5.5\,\%$ at $520\,$nm. \\

In Fig. \ref{fig:spectral_QE} we show the spectral response measurement of TWH013. One day after growth a QE$\,=3.7\,\%$ was measured at $520\,$nm ($2.38$\,eV). An XPS measurement of the photocathode was done, resulting in a further decrease of the QE. Afterwards, the cathode was moved into the vacuum suitcase and transferred to PhoTEx.

\subsection{Characterization at PhoTEx}
\begin{figure}
    \centering
    \includegraphics[width=\linewidth]{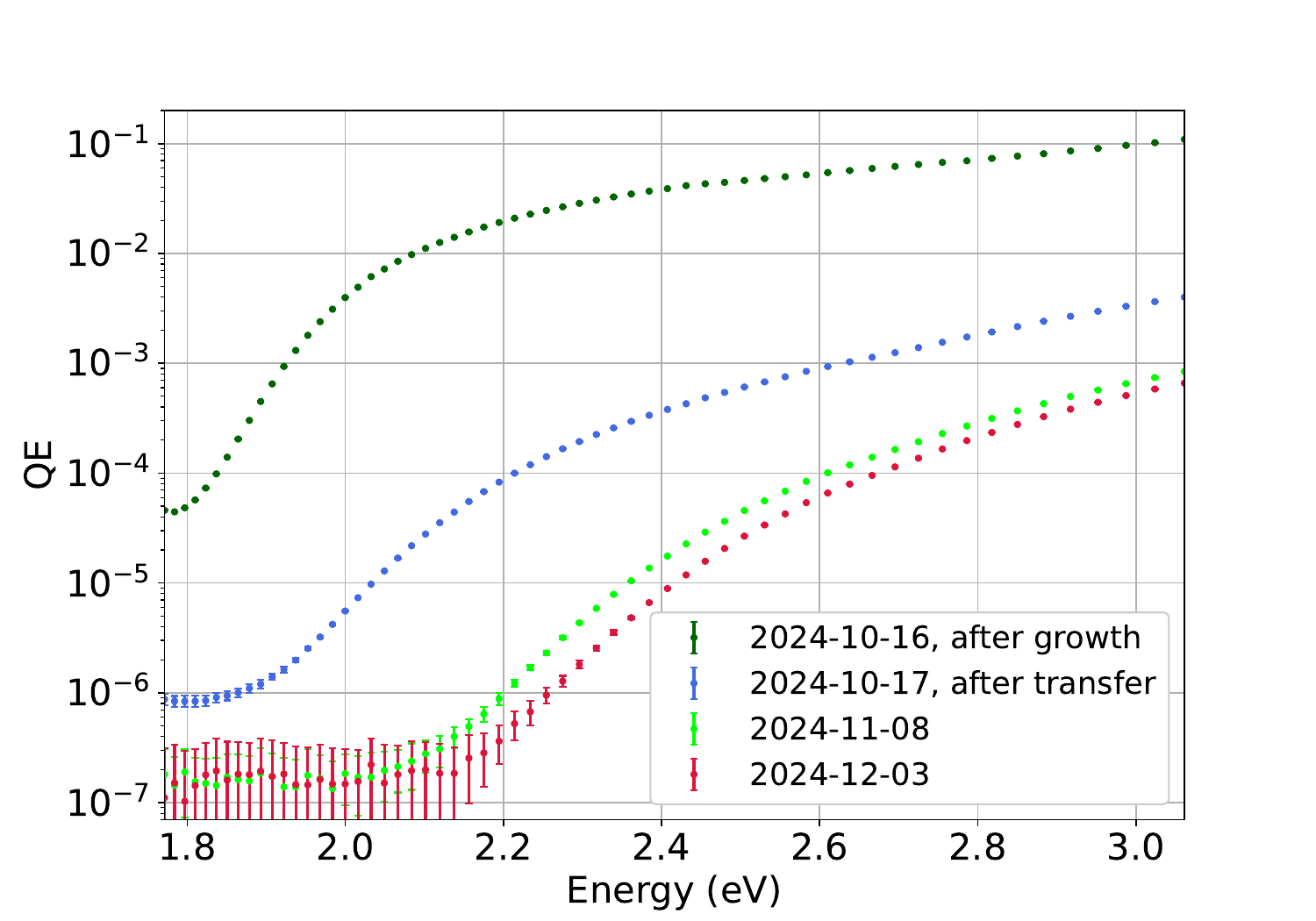}
    \caption{Spectral response curve of the photocathode TWH013. The first curve was measured in the preparation system. The other curves were measured at PhoTEx.}
    \label{fig:spectral_QE}
\end{figure}

The photocathode TWH013 was transferred to PhoTEx for the commissioning of the instrument. The spectral response of the photocathode was regularly measured and is shown in Fig. \ref{fig:spectral_QE}. The first curve was measured in the preparation chamber of the photocathode laboratory one day after the growth on October 16, 2024. An increased pressure during the transfer led to a significant reduction of the quantum efficiency. This happened due to pressure bursts up to high $10^{-9}\,$mbar when moving the transfer arm and led to a remaining QE of just $0.04\,\%$ at $2.38\,$eV. Note that the recipe was not yet optimized for a long lifetime. The recipe will be optimized in the future and the vacuum suitcase and load locks will be improved to optimize the vacuum conditions during the transfer.

Due to the excellent vacuum conditions at PhoTEx the photocathode degraded more slowly and there was still a detectable photocurrent after 7 weeks. Together with the decrease in QE, the effective workfunction appears to shift towards higher energies. Nevertheless, the high sensitivity of PhoTEx allows the measurement of the quantum efficiency down to $QE = 1\times 10^{-6}$. \\

\begin{figure}
    \centering
    \includegraphics[width=\linewidth]{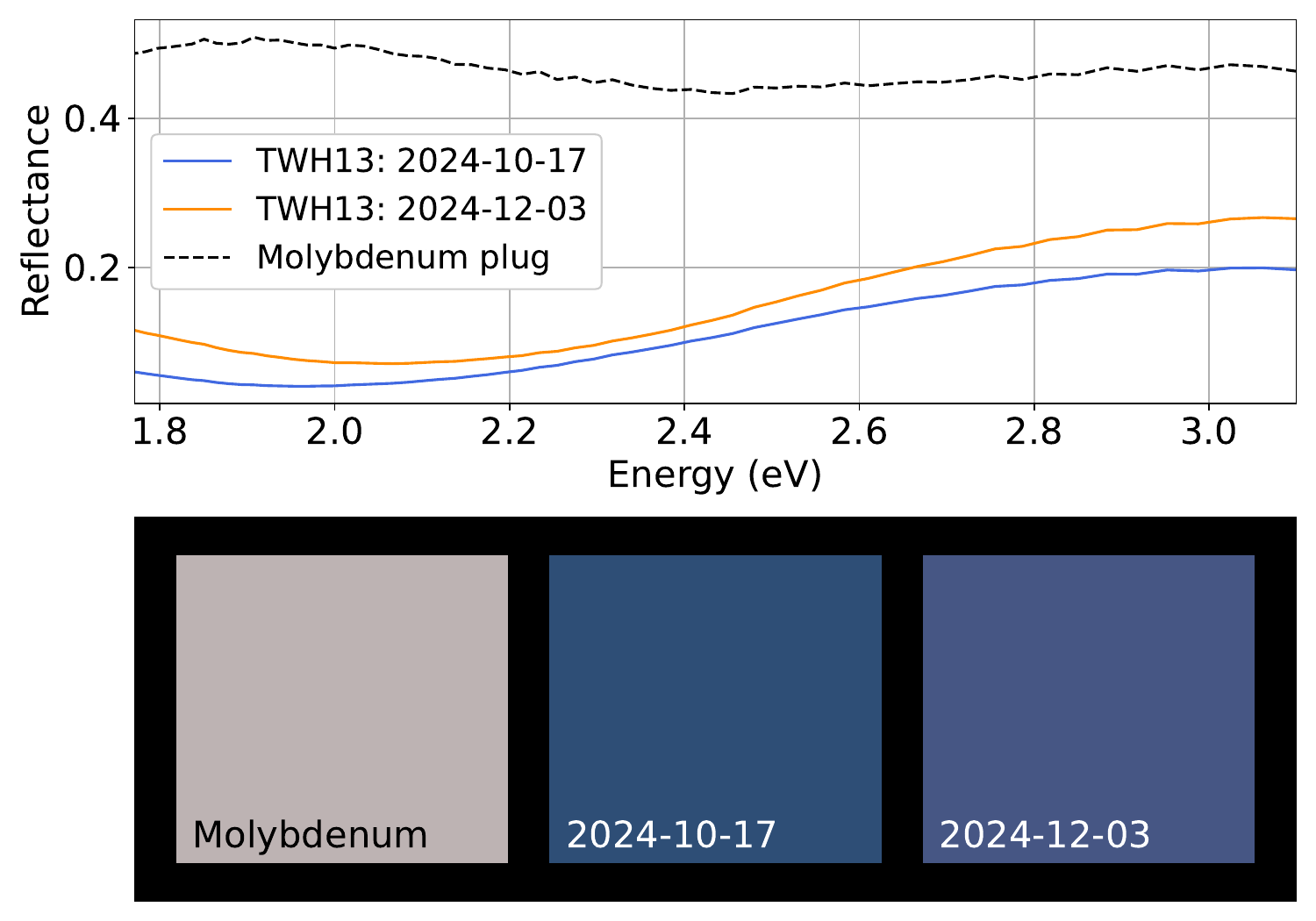}
    \caption{Reflectance curves of the photocathode TWH013. From the reflectance, the color was determined using the CIE 1931 color space. The reflectance and color of the pure molybdenum plug was measured as a reference.}
    \label{fig:Refl_color}
\end{figure}

Along with the spectral response, the reflectance was measured for an angle of incidence of $38.7^\circ$. As the reflectance changes less than the QE, only a selection of the measurements is shown in Fig. \ref{fig:Refl_color}. From the reflectance curves the color of the photocathode was determined as described above. The photocathode becomes slightly brighter and more purple as it ages. The reflectance of the pure molybdenum at the edge of the plug was also measured as a reference. The XYZ and sRGB values are listed in Tab. \ref{tab:TWH13}.

\begin{table}
\caption{Colorimetry data of the photocathode TWH013 measured at $38.7^\circ$ angle of incidence.}
\label{tab:TWH13}
\begin{ruledtabular}
\begin{tabular}{r l @{\hspace{8em}}}
    2024-10-17 & XYZ = 0.071; 0.073; 0.180 \\
     & sRGB = 45; 77; 117 \\
    2024-12-03 & XYZ =  0.100; 0.097; 0.231 \\
     & sRGB = 69; 86; 131 \\
     Molybdenum & XYZ = 0.452; 0.463; 0.492 \\
      & sRGB = 188; 178; 179 \\
\end{tabular}
\end{ruledtabular}
\end{table}

\begin{figure}
    \centering
    \includegraphics[width=\linewidth]{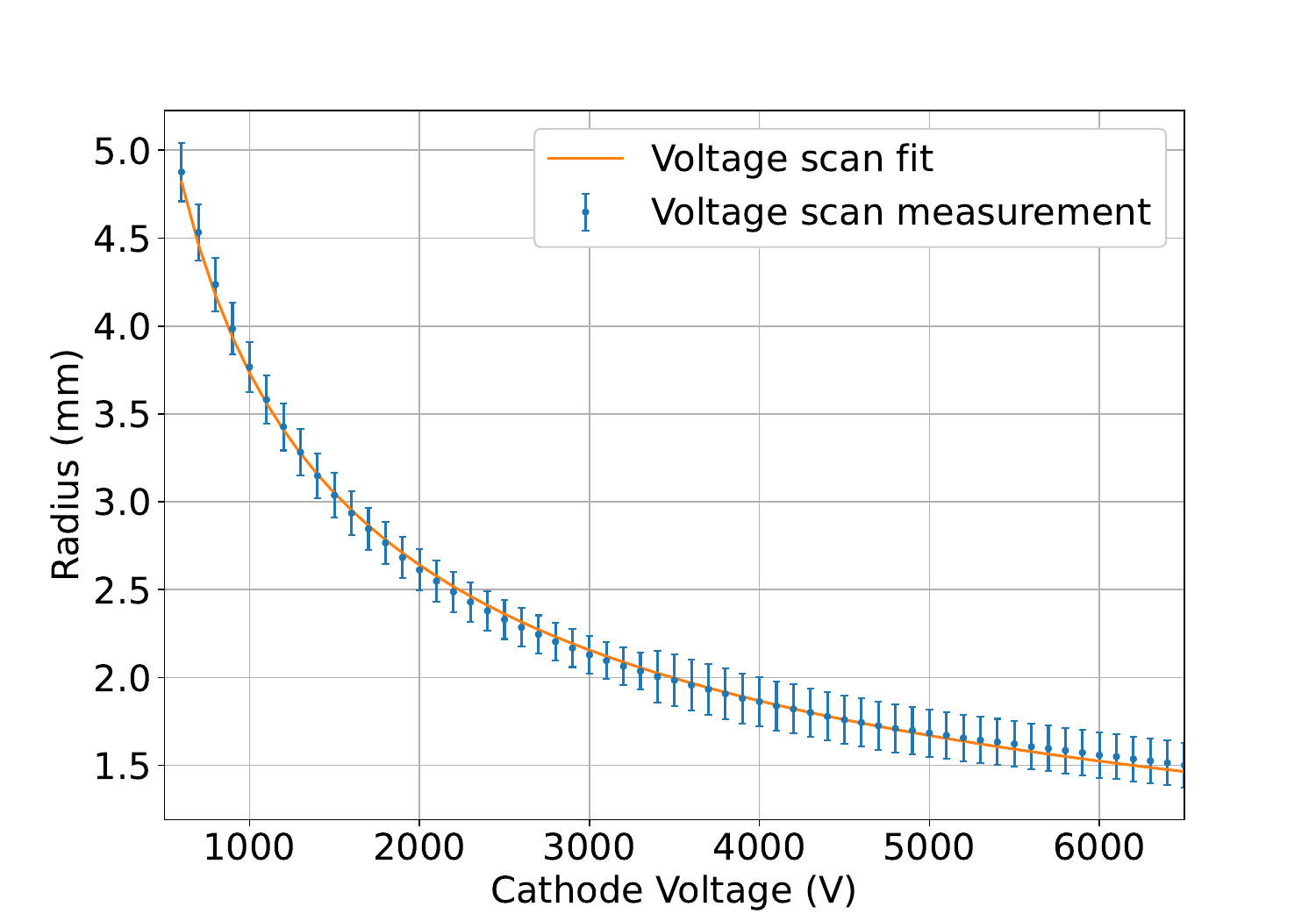}
    \caption{Voltage scan of the photocathode TWH013 to measure the MTE. The electron beam radius at the detector is measured as a function of the bias voltage at the cathode and fitted with Eq. \ref{eq:voltage_scan}. From the fit parameter, the MTE can be determined to be MTE $ =(135.5\pm2.5)\,$meV.}
    \label{fig:voltage_scan}
\end{figure}

The MTE was measured 67 days after the growth of the photocathode, on December 22, 2024. A first voltage scan ranging from $(600-6500)\,$V with steps of $100\,$V was performed at $450\,$nm ($2.76\,$eV). The voltage of the MCP was adjusted during the scan to ensure uniform brightness of the images with a fixed exposure time of $1\,$s. The Helmholtz coils were used to steer the beam to the center of the detector. The scan is shown in Fig. \ref{fig:voltage_scan} and results in a measured value of MTE$\,=(135.5\pm2.5)\,$meV. This corresponds to a relative uncertainty of $2\,\%$. 
Each measured point was also analyzed as a single shot measurement using Eq. \ref{eq:MTE}. Both methods agree with each other within the limits of uncertainty. Anyhow, the uncertainties of the single shot measurements are significantly larger, ranging from $7\,\%$ up to $17\,\%$ for small electron beam radii at the detector.
An optical power of $28\,$nW was used for this measurement. Together with the known QE, the photocurrent $I$ and average number of electrons in the system $N_e$ can be determined with Eq. \ref{eq:number_electrons}. With a current of $I = 2\times 10^{-11}\,$A and $N_e=1.5$ we do not expect any space charge effects. In the case of a photocathode with higher QE, the optical power should be attenuated using ND filters.

\begin{figure}
    \centering
    \includegraphics[width=\linewidth]{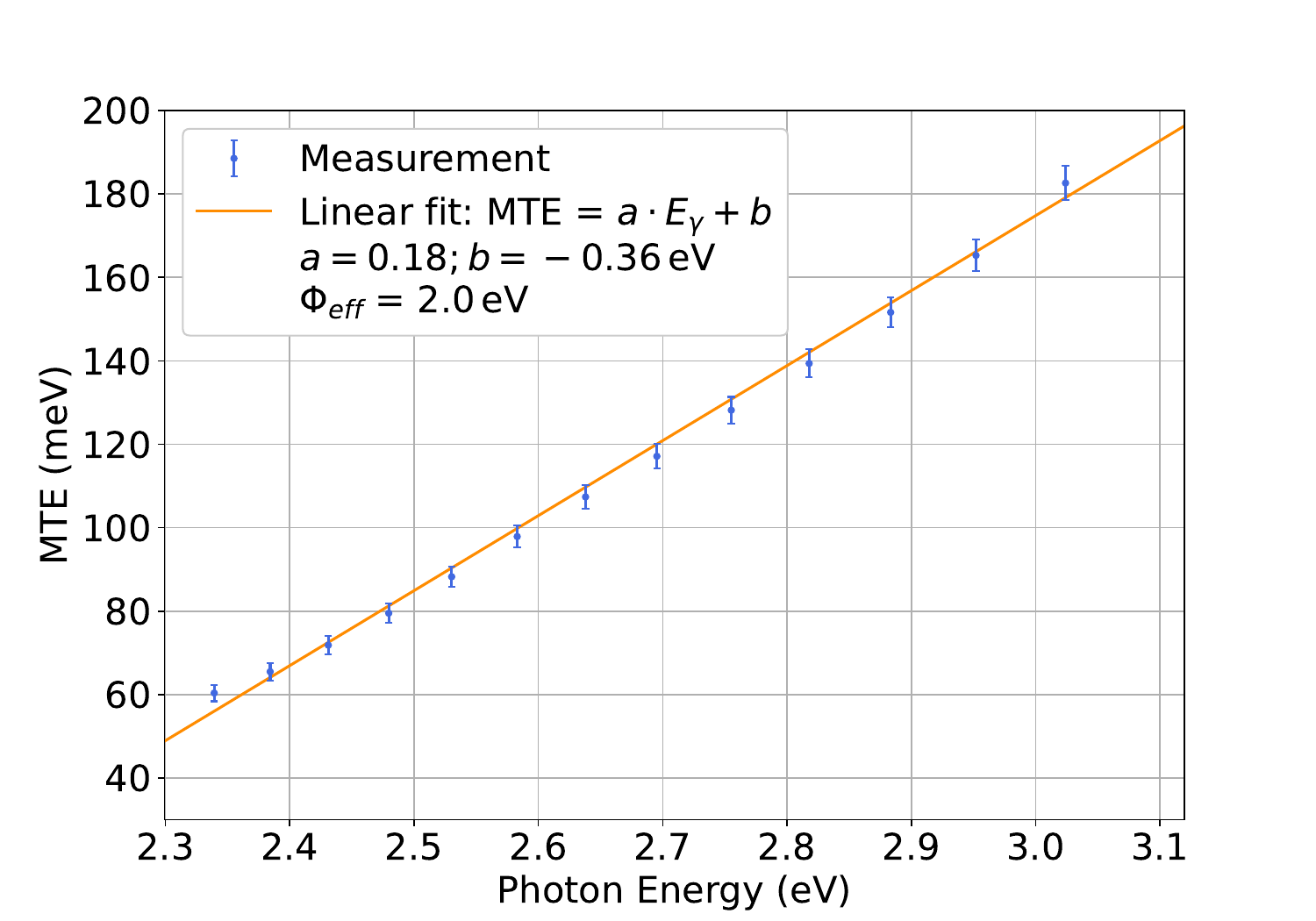}
    \caption{Wavelength scan of photocathode TWH013. The data can be fitted with a linear function to determine the effective workfunction $\phi_\text{eff} = 2.0\,$eV.}
    \label{fig:Wvl_scan}
\end{figure}

Additionally, the MTE was measured as a function of the wavelength from $400\,$nm to $530\,$nm. For longer wavelengths, a full voltage scan is no longer possible due to the low remaining quantum efficiency of the photocathode. Therefore, PhoTEx is able to measure the MTE up to a QE of about $0.001\,\%$. For each wavelength a voltage scan from $(700-6500)\,$V with a step width of $200\,$V was performed to determine the MTE. The data is plotted in Fig. \ref{fig:Wvl_scan} as a function of the photon energy. This presentation allows an easy determination of the effective workfunction by fitting the data with a linear function. The effective workfunction $\phi_\text{eff}$ is the barrier, the electron has to overcome in order to get emitted.\cite{saha2023} In a first order approximation we expect MTE$\,=0\,$meV when the excitation energy is equal $\phi_\text{eff}$ as stated in Eq. \ref{eq:QE_MTE_theory}. From the linear fit we obtain $\phi_\text{eff}=2.0\,$eV.

Maxson \textit{et al.} measured the MTE of a sequentially deposited Na-K-Sb photocathode on a stainless steel substrate to be $134\,$meV at a wavelength of $532\,$nm.\cite{maxson2015} The results presented here are significantly smaller. However, the TWH013 photocathode was measured at an aged state. As shown in Fig. \ref{fig:spectral_QE} the aging leads to a shift in the work function towards higher energies and therefore reducing the MTE. In the range considered by the wavelength scan, the MTE is already tending towards the thermal limit. The slope of the linear function is therefore smaller then the expected value $a=1/3$ which has already been observed in earlier studies and the calculated effective workfunction is underestimated.\cite{kachwala2023} This has to be compared with other, less aged photocathodes in the future.

\section{Conclusion}

A new preparation chamber was set up at the photocathode laboratory at HZB. The system was upgraded with effusion cells for the alkali metals and a thermal cracker cell for antimony. As a result, more photocathodes can be grown without changing the sources, which improves the base pressure of the vacuum system and the reproducibility of the photocathode recipes as well as the time required for the preparation and growth of the photocathodes. All three sources are now pointed at the same position, allowing photocathode growth using the triple evaporation method, where all three materials are deposited simultaneously. \\

The novel characterization system PhoTEx is able to measure the MTE, QE, reflectance and lifetime of the photocathodes. PhoTEx is an instrument allowing the measurement of all physical key parameters in one compact setup without moving the sample. This is of major importance, as any transfer to another system may harm the photocathode.
PhoTEx allows characterization of the photocathodes in the full visible regime. The spectral resolved measurement of the reflectance enables the study of a new parameter: the color of the photocathode. The color has already been subject of much discussion in the photocathode community, to allow an easy first assessment of the photocathode quality. For the first time, the photocathode colorimetry allows a systematic analysis of this parameter. The high resolution of PhoTEx allows the characterization over a long period of time. The QE can be measured down to QE $=1\times 10^{-6}$ and the MTE up to an efficiency of at least QE $=1\times 10^{-5}$. The MTE can be measured by a single shot measurement or a voltage scan. Since the voltage scan consists of several measurements, the uncertainty is significantly smaller with this method. \\

After commissioning, a triple evaporation photocathode was grown with the new preparation system. Best to our knowledge, we give the first detailed description of the Na-K-Sb triple evaporation growth procedure. Directly after growth the photocathode reached a QE of $5.5\,\%$ at $520\,$nm. The photocathode was transferred to PhoTEx and analyzed for more then 2 month. The degradation of the quantum efficiency, reflectance and colorimetry was studied at PhoTEx. The MTE was measured with two different methods and both results agree with each other within the limits of uncertainty. Overall, all measurements are self-consistent and provide a credible dataset. \\

In the future, PhoTEx will be used to routinely study the photocathodes grown at HZB regarding their QE, MTE and lifetime. This will help to further improve the photocathode recipes at the new triple evaporation growth system. Together they form the basis of new photocathodes with high QE and low MTE which are crucial for the generation of high-brightness electron beams at existing and future accelerator facilities.

\section*{Acknowledgements}
The authors gratefully acknowledge the support from Axel Neumann and Jens Völker from Helmholtz-Zentrum Berlin as well as the consultancy and scientific exchange with Laura Monaco and Daniele Sertore from INFN Milano - LASA and Lee Jones from STFC Daresbury Laboratory.

Work supported by German Bundesministerium für Bildung und Forschung, Land Berlin, grants of Helmholtz As-
sociation, and Deutsche Forschungsgemeinschaft (DFG): CO 1509/10-1 | MI 2917/1-1.

\section*{Author Declarations}
\subsection*{Conflict of Interest}
The authors have no conflicts to disclose.

\subsection*{Author Contributions}
\textbf{J. Dube:} PhoTEx - conceptualization (lead); Investigation (lead); Data curation (lead); Formal analysis (lead); Writing - original draft (lead); Visualization (lead); Writing - review \& editing (equal).
\textbf{J. Kühn:} Supervision (equal); Project administration (equal); Preparation chamber - conceptualization (lead); Investigation (equal); Writing - original draft (equal); Writing - review \& editing (equal).
\textbf{C. Wang:} Investigation (equal); Data curation (equal); Writing- review \& editing (equal).
\textbf{S. Mistry:} Funding acquisition (equal); Investigation (supporting); Writing - review \& editing (equal).
\textbf{G. Klemz:} Investigation (supporting); Writing- review \& editing (equal).
\textbf{A. Galdi:} Investigation (equal); Writing - review \& editing (equal).
\textbf{T. Kamps:} Supervision (equal); Funding acquisition (lead); Project administration (equal); Writing - review \& editing (equal).

\section*{Data Availability}
The data that support the findings of this study are available from the corresponding author upon reasonable request.

\section*{References}

\bibliography{PhoTEx_Paper}% Produces the bibliography via BibTeX.

%merlin.mbs aipnum4-1.bst 2010-07-25 4.21a (PWD, AO, DPC) hacked
%Control: key (0)
%Control: author (8) initials jnrlst
%Control: editor formatted (1) identically to author
%Control: production of article title (0) allowed
%Control: page (1) range
%Control: year (1) truncated
%Control: production of eprint (0) enabled
\begin{thebibliography}{37}%
\makeatletter
\providecommand \@ifxundefined [1]{%
 \@ifx{#1\undefined}
}%
\providecommand \@ifnum [1]{%
 \ifnum #1\expandafter \@firstoftwo
 \else \expandafter \@secondoftwo
 \fi
}%
\providecommand \@ifx [1]{%
 \ifx #1\expandafter \@firstoftwo
 \else \expandafter \@secondoftwo
 \fi
}%
\providecommand \natexlab [1]{#1}%
\providecommand \enquote  [1]{``#1''}%
\providecommand \bibnamefont  [1]{#1}%
\providecommand \bibfnamefont [1]{#1}%
\providecommand \citenamefont [1]{#1}%
\providecommand \href@noop [0]{\@secondoftwo}%
\providecommand \href [0]{\begingroup \@sanitize@url \@href}%
\providecommand \@href[1]{\@@startlink{#1}\@@href}%
\providecommand \@@href[1]{\endgroup#1\@@endlink}%
\providecommand \@sanitize@url [0]{\catcode `\\12\catcode `\$12\catcode `\&12\catcode `\#12\catcode `\^12\catcode `\_12\catcode `\%12\relax}%
\providecommand \@@startlink[1]{}%
\providecommand \@@endlink[0]{}%
\providecommand \url  [0]{\begingroup\@sanitize@url \@url }%
\providecommand \@url [1]{\endgroup\@href {#1}{\urlprefix }}%
\providecommand \urlprefix  [0]{URL }%
\providecommand \Eprint [0]{\href }%
\providecommand \doibase [0]{http://dx.doi.org/}%
\providecommand \selectlanguage [0]{\@gobble}%
\providecommand \bibinfo  [0]{\@secondoftwo}%
\providecommand \bibfield  [0]{\@secondoftwo}%
\providecommand \translation [1]{[#1]}%
\providecommand \BibitemOpen [0]{}%
\providecommand \bibitemStop [0]{}%
\providecommand \bibitemNoStop [0]{.\EOS\space}%
\providecommand \EOS [0]{\spacefactor3000\relax}%
\providecommand \BibitemShut  [1]{\csname bibitem#1\endcsname}%
\let\auto@bib@innerbib\@empty
%</preamble>
\bibitem [{\citenamefont {Zhao}\ \emph {et~al.}(2021)\citenamefont {Zhao}, \citenamefont {Wang}, \citenamefont {Feng}, \citenamefont {Chen},\ and\ \citenamefont {Cao}}]{zhao2021}%
  \BibitemOpen
  \bibfield  {author} {\bibinfo {author} {\bibfnamefont {Z.~T.}\ \bibnamefont {Zhao}}, \bibinfo {author} {\bibfnamefont {Z.}~\bibnamefont {Wang}}, \bibinfo {author} {\bibfnamefont {C.}~\bibnamefont {Feng}}, \bibinfo {author} {\bibfnamefont {S.}~\bibnamefont {Chen}}, \ and\ \bibinfo {author} {\bibfnamefont {L.}~\bibnamefont {Cao}},\ }\bibfield  {title} {\enquote {\bibinfo {title} {Energy recovery linac based fully coherent light source},}\ }\href {\doibase 10.1038/s41598-021-03354-0} {\bibfield  {journal} {\bibinfo  {journal} {Scientific Reports}\ }\textbf {\bibinfo {volume} {11}},\ \bibinfo {pages} {23875} (\bibinfo {year} {2021})}\BibitemShut {NoStop}%
\bibitem [{\citenamefont {Neumann}\ \emph {et~al.}(2022)\citenamefont {Neumann}, \citenamefont {Alberdi}, \citenamefont {Birke}, \citenamefont {Echevarria}, \citenamefont {Eichel}, \citenamefont {Falkenstern}, \citenamefont {Fleischhauer}, \citenamefont {Frahm}, \citenamefont {Goebel}, \citenamefont {Heugel} \emph {et~al.}}]{neumann2022}%
  \BibitemOpen
  \bibfield  {author} {\bibinfo {author} {\bibfnamefont {A.}~\bibnamefont {Neumann}}, \bibinfo {author} {\bibfnamefont {B.}~\bibnamefont {Alberdi}}, \bibinfo {author} {\bibfnamefont {T.}~\bibnamefont {Birke}}, \bibinfo {author} {\bibfnamefont {P.}~\bibnamefont {Echevarria}}, \bibinfo {author} {\bibfnamefont {D.}~\bibnamefont {Eichel}}, \bibinfo {author} {\bibfnamefont {F.}~\bibnamefont {Falkenstern}}, \bibinfo {author} {\bibfnamefont {R.}~\bibnamefont {Fleischhauer}}, \bibinfo {author} {\bibfnamefont {A.}~\bibnamefont {Frahm}}, \bibinfo {author} {\bibfnamefont {F.}~\bibnamefont {Goebel}}, \bibinfo {author} {\bibfnamefont {A.}~\bibnamefont {Heugel}},  \emph {et~al.},\ }\bibfield  {title} {\enquote {\bibinfo {title} {{bERLinPro becomes SEALab: Status and perspective of the energy recovery linac at HZB}},}\ }in\ \href {\doibase 18429/JACoW-IPAC2022-TUPOPT048} {\emph {\bibinfo {booktitle} {Proc. IPAC'22}}}\ (\bibinfo {year} {2022})\ pp.\ \bibinfo {pages} {1110--1113}\BibitemShut {NoStop}%
\bibitem [{\citenamefont {Emma}\ \emph {et~al.}(2010)\citenamefont {Emma}, \citenamefont {Akre}, \citenamefont {Arthur}, \citenamefont {Bionta}, \citenamefont {Bostedt}, \citenamefont {Bozek}, \citenamefont {Brachmann}, \citenamefont {Bucksbaum}, \citenamefont {Coffee}, \citenamefont {Decker} \emph {et~al.}}]{FEL_SLAC}%
  \BibitemOpen
  \bibfield  {author} {\bibinfo {author} {\bibfnamefont {P.}~\bibnamefont {Emma}}, \bibinfo {author} {\bibfnamefont {R.}~\bibnamefont {Akre}}, \bibinfo {author} {\bibfnamefont {J.}~\bibnamefont {Arthur}}, \bibinfo {author} {\bibfnamefont {R.}~\bibnamefont {Bionta}}, \bibinfo {author} {\bibfnamefont {C.}~\bibnamefont {Bostedt}}, \bibinfo {author} {\bibfnamefont {J.}~\bibnamefont {Bozek}}, \bibinfo {author} {\bibfnamefont {A.}~\bibnamefont {Brachmann}}, \bibinfo {author} {\bibfnamefont {P.}~\bibnamefont {Bucksbaum}}, \bibinfo {author} {\bibfnamefont {R.}~\bibnamefont {Coffee}}, \bibinfo {author} {\bibfnamefont {F.-J.}\ \bibnamefont {Decker}},  \emph {et~al.},\ }\bibfield  {title} {\enquote {\bibinfo {title} {First lasing and operation of an {\aa}ngstrom-wavelength free-electron laser},}\ }\href {\doibase 10.1038/nphoton.2010.176} {\bibfield  {journal} {\bibinfo  {journal} {nature photonics}\ }\textbf {\bibinfo {volume} {4}},\ \bibinfo {pages} {641--647} (\bibinfo {year} {2010})}\BibitemShut {NoStop}%
\bibitem [{\citenamefont {Schaber}, \citenamefont {Xiang},\ and\ \citenamefont {Gaponik}(2023)}]{schaber2023}%
  \BibitemOpen
  \bibfield  {author} {\bibinfo {author} {\bibfnamefont {J.}~\bibnamefont {Schaber}}, \bibinfo {author} {\bibfnamefont {R.}~\bibnamefont {Xiang}}, \ and\ \bibinfo {author} {\bibfnamefont {N.}~\bibnamefont {Gaponik}},\ }\bibfield  {title} {\enquote {\bibinfo {title} {Review of photocathodes for electron beam sources in particle accelerators},}\ }\href {\doibase 10.1039/D2TC03729G} {\bibfield  {journal} {\bibinfo  {journal} {Journal of Materials Chemistry C}\ }\textbf {\bibinfo {volume} {11}},\ \bibinfo {pages} {3162--3179} (\bibinfo {year} {2023})}\BibitemShut {NoStop}%
\bibitem [{\citenamefont {Filippetto}\ \emph {et~al.}(2022)\citenamefont {Filippetto}, \citenamefont {Musumeci}, \citenamefont {Li}, \citenamefont {Siwick}, \citenamefont {Otto}, \citenamefont {Centurion},\ and\ \citenamefont {Nunes}}]{filippetto2022}%
  \BibitemOpen
  \bibfield  {author} {\bibinfo {author} {\bibfnamefont {D.}~\bibnamefont {Filippetto}}, \bibinfo {author} {\bibfnamefont {P.}~\bibnamefont {Musumeci}}, \bibinfo {author} {\bibfnamefont {R.}~\bibnamefont {Li}}, \bibinfo {author} {\bibfnamefont {B.}~\bibnamefont {Siwick}}, \bibinfo {author} {\bibfnamefont {M.}~\bibnamefont {Otto}}, \bibinfo {author} {\bibfnamefont {M.}~\bibnamefont {Centurion}}, \ and\ \bibinfo {author} {\bibfnamefont {J.}~\bibnamefont {Nunes}},\ }\bibfield  {title} {\enquote {\bibinfo {title} {Ultrafast electron diffraction: {Visualizing} dynamic states of matter},}\ }\href {\doibase 10.1103/RevModPhys.94.045004} {\bibfield  {journal} {\bibinfo  {journal} {Reviews of Modern Physics}\ }\textbf {\bibinfo {volume} {94}},\ \bibinfo {pages} {045004} (\bibinfo {year} {2022})}\BibitemShut {NoStop}%
\bibitem [{\citenamefont {Zewail}(2006)}]{zewail2006}%
  \BibitemOpen
  \bibfield  {author} {\bibinfo {author} {\bibfnamefont {A.~H.}\ \bibnamefont {Zewail}},\ }\bibfield  {title} {\enquote {\bibinfo {title} {{4D} {Ultrafast} {Electron} {Diffraction}, {Crystallography}, {and} {Microscopy}},}\ }\href {\doibase 10.1146/annurev.physchem.57.032905.104748} {\bibfield  {journal} {\bibinfo  {journal} {Annual Review of Physical Chemistry}\ }\textbf {\bibinfo {volume} {57}},\ \bibinfo {pages} {65--103} (\bibinfo {year} {2006})}\BibitemShut {NoStop}%
\bibitem [{\citenamefont {Alberdi~Esuain}\ \emph {et~al.}(2022)\citenamefont {Alberdi~Esuain}, \citenamefont {Hwang}, \citenamefont {Neumann},\ and\ \citenamefont {Kamps}}]{Esuain2022_UED}%
  \BibitemOpen
  \bibfield  {author} {\bibinfo {author} {\bibfnamefont {B.}~\bibnamefont {Alberdi~Esuain}}, \bibinfo {author} {\bibfnamefont {J.-G.}\ \bibnamefont {Hwang}}, \bibinfo {author} {\bibfnamefont {A.}~\bibnamefont {Neumann}}, \ and\ \bibinfo {author} {\bibfnamefont {T.}~\bibnamefont {Kamps}},\ }\bibfield  {title} {\enquote {\bibinfo {title} {Novel approach to push the limit of temporal resolution in ultrafast electron diffraction accelerators},}\ }\href {\doibase 10.1038/s41598-022-17453-z} {\bibfield  {journal} {\bibinfo  {journal} {Scientific Reports}\ }\textbf {\bibinfo {volume} {12}},\ \bibinfo {pages} {13365} (\bibinfo {year} {2022})}\BibitemShut {NoStop}%
\bibitem [{\citenamefont {Rao}\ and\ \citenamefont {Dowell}(2014)}]{rao2014}%
  \BibitemOpen
  \bibfield  {author} {\bibinfo {author} {\bibfnamefont {T.}~\bibnamefont {Rao}}\ and\ \bibinfo {author} {\bibfnamefont {D.~H.}\ \bibnamefont {Dowell}},\ }\href@noop {} {\emph {\bibinfo {title} {An Engineering Guide To Photoinjectors}}}\ (\bibinfo  {publisher} {arXiv.1403.7539},\ \bibinfo {year} {2014})\ \Eprint {http://arxiv.org/abs/1403.7539} {1403.7539} \BibitemShut {NoStop}%
\bibitem [{\citenamefont {Lee}\ \emph {et~al.}(2015)\citenamefont {Lee}, \citenamefont {Karkare}, \citenamefont {Cultrera}, \citenamefont {Kim},\ and\ \citenamefont {Bazarov}}]{Lee2015}%
  \BibitemOpen
  \bibfield  {author} {\bibinfo {author} {\bibfnamefont {H.}~\bibnamefont {Lee}}, \bibinfo {author} {\bibfnamefont {S.}~\bibnamefont {Karkare}}, \bibinfo {author} {\bibfnamefont {L.}~\bibnamefont {Cultrera}}, \bibinfo {author} {\bibfnamefont {A.}~\bibnamefont {Kim}}, \ and\ \bibinfo {author} {\bibfnamefont {I.~V.}\ \bibnamefont {Bazarov}},\ }\bibfield  {title} {\enquote {\bibinfo {title} {Review and demonstration of ultra-low-emittance photocathode measurements},}\ }\href {\doibase 10.1063/1.4927381} {\bibfield  {journal} {\bibinfo  {journal} {Review of Scientific Instruments}\ }\textbf {\bibinfo {volume} {86}},\ \bibinfo {pages} {073309} (\bibinfo {year} {2015})}\BibitemShut {NoStop}%
\bibitem [{\citenamefont {Schmüser}\ \emph {et~al.}(2014)\citenamefont {Schmüser}, \citenamefont {Dohlus}, \citenamefont {Rossbach}, \citenamefont {Behrens}, \citenamefont {Behrens}, \citenamefont {Dohlus},\ and\ \citenamefont {Rossbach}}]{Schmueser2014}%
  \BibitemOpen
  \bibfield  {author} {\bibinfo {author} {\bibfnamefont {P.}~\bibnamefont {Schmüser}}, \bibinfo {author} {\bibfnamefont {M.}~\bibnamefont {Dohlus}}, \bibinfo {author} {\bibfnamefont {J.}~\bibnamefont {Rossbach}}, \bibinfo {author} {\bibfnamefont {C.}~\bibnamefont {Behrens}}, \bibinfo {author} {\bibfnamefont {C.}~\bibnamefont {Behrens}}, \bibinfo {author} {\bibfnamefont {M.}~\bibnamefont {Dohlus}}, \ and\ \bibinfo {author} {\bibfnamefont {J.}~\bibnamefont {Rossbach}},\ }\href@noop {} {\emph {\bibinfo {title} {Free-Electron Lasers in the Ultraviolet and X-Ray Regime: Physical Principles, Experimental Results, Technical Realization}}},\ \bibinfo {edition} {2nd}\ ed.,\ \bibinfo {series} {Springer Tracts in Modern Physics}, Vol.\ \bibinfo {volume} {258}\ (\bibinfo  {publisher} {Springer Nature},\ \bibinfo {year} {2014})\BibitemShut {NoStop}%
\bibitem [{\citenamefont {Saha}\ \emph {et~al.}(2023)\citenamefont {Saha}, \citenamefont {Chubenko}, \citenamefont {Kevin~Nangoi}, \citenamefont {Arias}, \citenamefont {Montgomery}, \citenamefont {Poddar}, \citenamefont {Padmore},\ and\ \citenamefont {Karkare}}]{saha2023}%
  \BibitemOpen
  \bibfield  {author} {\bibinfo {author} {\bibfnamefont {P.}~\bibnamefont {Saha}}, \bibinfo {author} {\bibfnamefont {O.}~\bibnamefont {Chubenko}}, \bibinfo {author} {\bibfnamefont {J.}~\bibnamefont {Kevin~Nangoi}}, \bibinfo {author} {\bibfnamefont {T.}~\bibnamefont {Arias}}, \bibinfo {author} {\bibfnamefont {E.}~\bibnamefont {Montgomery}}, \bibinfo {author} {\bibfnamefont {S.}~\bibnamefont {Poddar}}, \bibinfo {author} {\bibfnamefont {H.~A.}\ \bibnamefont {Padmore}}, \ and\ \bibinfo {author} {\bibfnamefont {S.}~\bibnamefont {Karkare}},\ }\bibfield  {title} {\enquote {\bibinfo {title} {Theory of photoemission from cathodes with disordered surfaces},}\ }\href {\doibase 10.1063/5.0135629} {\bibfield  {journal} {\bibinfo  {journal} {Journal of Applied Physics}\ }\textbf {\bibinfo {volume} {133}} (\bibinfo {year} {2023}),\ 10.1063/5.0135629}\BibitemShut {NoStop}%
\bibitem [{\citenamefont {Dowell}\ \emph {et~al.}(2010)\citenamefont {Dowell}, \citenamefont {Bazarov}, \citenamefont {Dunham}, \citenamefont {Harkay}, \citenamefont {Hernandez-Garcia}, \citenamefont {Legg}, \citenamefont {Padmore}, \citenamefont {Rao}, \citenamefont {Smedley},\ and\ \citenamefont {Wan}}]{dowell2010}%
  \BibitemOpen
  \bibfield  {author} {\bibinfo {author} {\bibfnamefont {D.~H.}\ \bibnamefont {Dowell}}, \bibinfo {author} {\bibfnamefont {I.}~\bibnamefont {Bazarov}}, \bibinfo {author} {\bibfnamefont {B.}~\bibnamefont {Dunham}}, \bibinfo {author} {\bibfnamefont {K.}~\bibnamefont {Harkay}}, \bibinfo {author} {\bibfnamefont {C.}~\bibnamefont {Hernandez-Garcia}}, \bibinfo {author} {\bibfnamefont {R.}~\bibnamefont {Legg}}, \bibinfo {author} {\bibfnamefont {H.}~\bibnamefont {Padmore}}, \bibinfo {author} {\bibfnamefont {T.}~\bibnamefont {Rao}}, \bibinfo {author} {\bibfnamefont {J.}~\bibnamefont {Smedley}}, \ and\ \bibinfo {author} {\bibfnamefont {W.}~\bibnamefont {Wan}},\ }\bibfield  {title} {\enquote {\bibinfo {title} {Cathode {R}\&{D} for future light sources},}\ }\href {\doibase 10.1016/j.nima.2010.03.104} {\bibfield  {journal} {\bibinfo  {journal} {Nuclear Instruments and Methods in Physics Research Section A: Accelerators, Spectrometers, Detectors and Associated Equipment}\ }\textbf {\bibinfo {volume} {622}},\ \bibinfo
  {pages} {685--697} (\bibinfo {year} {2010})}\BibitemShut {NoStop}%
\bibitem [{\citenamefont {Wang}\ \emph {et~al.}(2017)\citenamefont {Wang}, \citenamefont {Pandey}, \citenamefont {Moody},\ and\ \citenamefont {Batista}}]{wang2017}%
  \BibitemOpen
  \bibfield  {author} {\bibinfo {author} {\bibfnamefont {G.}~\bibnamefont {Wang}}, \bibinfo {author} {\bibfnamefont {R.}~\bibnamefont {Pandey}}, \bibinfo {author} {\bibfnamefont {N.~A.}\ \bibnamefont {Moody}}, \ and\ \bibinfo {author} {\bibfnamefont {E.~R.}\ \bibnamefont {Batista}},\ }\bibfield  {title} {\enquote {\bibinfo {title} {Degradation of {Alkali}-{Based} {Photocathodes} from {Exposure} to {Residual} {Gases}: {A} {First}-{Principles} {Study}},}\ }\href {\doibase 10.1021/acs.jpcc.6b12796} {\bibfield  {journal} {\bibinfo  {journal} {The Journal of Physical Chemistry C}\ }\textbf {\bibinfo {volume} {121}},\ \bibinfo {pages} {8399--8408} (\bibinfo {year} {2017})}\BibitemShut {NoStop}%
\bibitem [{\citenamefont {Wang}\ \emph {et~al.}(2021)\citenamefont {Wang}, \citenamefont {Litvinenko}, \citenamefont {Pinayev}, \citenamefont {Gaowei}, \citenamefont {Skaritka}, \citenamefont {Belomestnykh}, \citenamefont {Ben-Zvi}, \citenamefont {Brutus}, \citenamefont {Jing}, \citenamefont {Biswas}, \citenamefont {Ma}, \citenamefont {Narayan}, \citenamefont {Petrushina}, \citenamefont {Rahman}, \citenamefont {Xin}, \citenamefont {Rao}, \citenamefont {Severino}, \citenamefont {Shih}, \citenamefont {Smith}, \citenamefont {Wang},\ and\ \citenamefont {Wu}}]{wang2021}%
  \BibitemOpen
  \bibfield  {author} {\bibinfo {author} {\bibfnamefont {E.}~\bibnamefont {Wang}}, \bibinfo {author} {\bibfnamefont {V.~N.}\ \bibnamefont {Litvinenko}}, \bibinfo {author} {\bibfnamefont {I.}~\bibnamefont {Pinayev}}, \bibinfo {author} {\bibfnamefont {M.}~\bibnamefont {Gaowei}}, \bibinfo {author} {\bibfnamefont {J.}~\bibnamefont {Skaritka}}, \bibinfo {author} {\bibfnamefont {S.}~\bibnamefont {Belomestnykh}}, \bibinfo {author} {\bibfnamefont {I.}~\bibnamefont {Ben-Zvi}}, \bibinfo {author} {\bibfnamefont {J.~C.}\ \bibnamefont {Brutus}}, \bibinfo {author} {\bibfnamefont {Y.}~\bibnamefont {Jing}}, \bibinfo {author} {\bibfnamefont {J.}~\bibnamefont {Biswas}}, \bibinfo {author} {\bibfnamefont {J.}~\bibnamefont {Ma}}, \bibinfo {author} {\bibfnamefont {G.}~\bibnamefont {Narayan}}, \bibinfo {author} {\bibfnamefont {I.}~\bibnamefont {Petrushina}}, \bibinfo {author} {\bibfnamefont {O.}~\bibnamefont {Rahman}}, \bibinfo {author} {\bibfnamefont {T.}~\bibnamefont {Xin}}, \bibinfo {author} {\bibfnamefont {T.}~\bibnamefont
  {Rao}}, \bibinfo {author} {\bibfnamefont {F.}~\bibnamefont {Severino}}, \bibinfo {author} {\bibfnamefont {K.}~\bibnamefont {Shih}}, \bibinfo {author} {\bibfnamefont {K.}~\bibnamefont {Smith}}, \bibinfo {author} {\bibfnamefont {G.}~\bibnamefont {Wang}}, \ and\ \bibinfo {author} {\bibfnamefont {Y.}~\bibnamefont {Wu}},\ }\bibfield  {title} {\enquote {\bibinfo {title} {Long lifetime of bialkali photocathodes operating in high gradient superconducting radio frequency gun},}\ }\href {\doibase 10.1038/s41598-021-83997-1} {\bibfield  {journal} {\bibinfo  {journal} {Scientific Reports}\ }\textbf {\bibinfo {volume} {11}},\ \bibinfo {pages} {4477} (\bibinfo {year} {2021})}\BibitemShut {NoStop}%
\bibitem [{\citenamefont {Schmeißer}\ \emph {et~al.}(2018)\citenamefont {Schmeißer}, \citenamefont {Mistry}, \citenamefont {Kirschner}, \citenamefont {Schubert}, \citenamefont {Jankowiak}, \citenamefont {Kamps},\ and\ \citenamefont {Kühn}}]{schmeisser2018}%
  \BibitemOpen
  \bibfield  {author} {\bibinfo {author} {\bibfnamefont {M.~A.}\ \bibnamefont {Schmeißer}}, \bibinfo {author} {\bibfnamefont {S.}~\bibnamefont {Mistry}}, \bibinfo {author} {\bibfnamefont {H.}~\bibnamefont {Kirschner}}, \bibinfo {author} {\bibfnamefont {S.}~\bibnamefont {Schubert}}, \bibinfo {author} {\bibfnamefont {A.}~\bibnamefont {Jankowiak}}, \bibinfo {author} {\bibfnamefont {T.}~\bibnamefont {Kamps}}, \ and\ \bibinfo {author} {\bibfnamefont {J.}~\bibnamefont {Kühn}},\ }\bibfield  {title} {\enquote {\bibinfo {title} {Towards the operation of {Cs}-{K}-{Sb} photocathodes in superconducting rf photoinjectors},}\ }\href {\doibase 10.1103/PhysRevAccelBeams.21.113401} {\bibfield  {journal} {\bibinfo  {journal} {Physical Review Accelerators and Beams}\ }\textbf {\bibinfo {volume} {21}},\ \bibinfo {pages} {113401} (\bibinfo {year} {2018})}\BibitemShut {NoStop}%
\bibitem [{\citenamefont {Cocchi}\ \emph {et~al.}(2019)\citenamefont {Cocchi}, \citenamefont {Mistry}, \citenamefont {Schmei{\ss}er}, \citenamefont {Amador}, \citenamefont {K{\"u}hn},\ and\ \citenamefont {Kamps}}]{cocchi2019}%
  \BibitemOpen
  \bibfield  {author} {\bibinfo {author} {\bibfnamefont {C.}~\bibnamefont {Cocchi}}, \bibinfo {author} {\bibfnamefont {S.}~\bibnamefont {Mistry}}, \bibinfo {author} {\bibfnamefont {M.}~\bibnamefont {Schmei{\ss}er}}, \bibinfo {author} {\bibfnamefont {R.}~\bibnamefont {Amador}}, \bibinfo {author} {\bibfnamefont {J.}~\bibnamefont {K{\"u}hn}}, \ and\ \bibinfo {author} {\bibfnamefont {T.}~\bibnamefont {Kamps}},\ }\bibfield  {title} {\enquote {\bibinfo {title} {Electronic structure and core electron fingerprints of caesium-based multi-alkali antimonides for ultra-bright electron sources},}\ }\href {\doibase 10.1038/s41598-019-54419-0} {\bibfield  {journal} {\bibinfo  {journal} {Scientific Reports}\ }\textbf {\bibinfo {volume} {9}},\ \bibinfo {pages} {18276} (\bibinfo {year} {2019})}\BibitemShut {NoStop}%
\bibitem [{\citenamefont {Mistry}\ \emph {et~al.}(2022)\citenamefont {Mistry}, \citenamefont {Kamps}, \citenamefont {K{\"u}hn},\ and\ \citenamefont {Wang}}]{mistry2022}%
  \BibitemOpen
  \bibfield  {author} {\bibinfo {author} {\bibfnamefont {S.}~\bibnamefont {Mistry}}, \bibinfo {author} {\bibfnamefont {T.}~\bibnamefont {Kamps}}, \bibinfo {author} {\bibfnamefont {J.}~\bibnamefont {K{\"u}hn}}, \ and\ \bibinfo {author} {\bibfnamefont {C.}~\bibnamefont {Wang}},\ }\bibfield  {title} {\enquote {\bibinfo {title} {Multi-alkali antimonide photocathode development for high brightness beams},}\ }in\ \href {\doibase 10.18429/JACoW-IPAC2022-THPOPT019} {\emph {\bibinfo {booktitle} {Proc. IPAC'22}}}\ (\bibinfo {year} {2022})\ pp.\ \bibinfo {pages} {2610--2613}\BibitemShut {NoStop}%
\bibitem [{\citenamefont {Mistry}\ \emph {et~al.}(2023)\citenamefont {Mistry}, \citenamefont {Dube}, \citenamefont {Kamps}, \citenamefont {Kuehn},\ and\ \citenamefont {Wang}}]{mistry2023}%
  \BibitemOpen
  \bibfield  {author} {\bibinfo {author} {\bibfnamefont {S.}~\bibnamefont {Mistry}}, \bibinfo {author} {\bibfnamefont {J.}~\bibnamefont {Dube}}, \bibinfo {author} {\bibfnamefont {T.}~\bibnamefont {Kamps}}, \bibinfo {author} {\bibfnamefont {J.}~\bibnamefont {Kuehn}}, \ and\ \bibinfo {author} {\bibfnamefont {C.}~\bibnamefont {Wang}},\ }\bibfield  {title} {\enquote {\bibinfo {title} {Evaluation of the in-situ photocathode handling for {SRF} photoinjector of {SEALab}},}\ }in\ \href {\doibase 10.18429/JACoW-IPAC2022-THPOPT019} {\emph {\bibinfo {booktitle} {Proc. IPAC'22}}}\ (\bibinfo {year} {2023})\ pp.\ \bibinfo {pages} {4240--4242},\ \bibinfo {note} {issue: 14}\BibitemShut {NoStop}%
\bibitem [{\citenamefont {Rozhkov}\ \emph {et~al.}(2024)\citenamefont {Rozhkov}, \citenamefont {Bakin}, \citenamefont {Rusetsky}, \citenamefont {Kustov}, \citenamefont {Golyashov}, \citenamefont {Demin}, \citenamefont {Scheibler}, \citenamefont {Alperovich},\ and\ \citenamefont {Tereshchenko}}]{rozhkov2024}%
  \BibitemOpen
  \bibfield  {author} {\bibinfo {author} {\bibfnamefont {S.}~\bibnamefont {Rozhkov}}, \bibinfo {author} {\bibfnamefont {V.}~\bibnamefont {Bakin}}, \bibinfo {author} {\bibfnamefont {V.}~\bibnamefont {Rusetsky}}, \bibinfo {author} {\bibfnamefont {D.}~\bibnamefont {Kustov}}, \bibinfo {author} {\bibfnamefont {V.}~\bibnamefont {Golyashov}}, \bibinfo {author} {\bibfnamefont {A.}~\bibnamefont {Demin}}, \bibinfo {author} {\bibfnamefont {H.}~\bibnamefont {Scheibler}}, \bibinfo {author} {\bibfnamefont {V.}~\bibnamefont {Alperovich}}, \ and\ \bibinfo {author} {\bibfnamefont {O.}~\bibnamefont {Tereshchenko}},\ }\bibfield  {title} {\enquote {\bibinfo {title} {Na 2 {KSb} / {Cs} x {Sb} interface engineering for high-efficiency photocathodes},}\ }\href {\doibase 10.1103/PhysRevApplied.22.024008} {\bibfield  {journal} {\bibinfo  {journal} {Physical Review Applied}\ }\textbf {\bibinfo {volume} {22}} (\bibinfo {year} {2024}),\ 10.1103/PhysRevApplied.22.024008}\BibitemShut {NoStop}%
\bibitem [{\citenamefont {Feng}\ \emph {et~al.}(2015)\citenamefont {Feng}, \citenamefont {Nasiatka}, \citenamefont {Wan}, \citenamefont {Vecchione},\ and\ \citenamefont {Padmore}}]{Feng2015}%
  \BibitemOpen
  \bibfield  {author} {\bibinfo {author} {\bibfnamefont {J.}~\bibnamefont {Feng}}, \bibinfo {author} {\bibfnamefont {J.}~\bibnamefont {Nasiatka}}, \bibinfo {author} {\bibfnamefont {W.}~\bibnamefont {Wan}}, \bibinfo {author} {\bibfnamefont {T.}~\bibnamefont {Vecchione}}, \ and\ \bibinfo {author} {\bibfnamefont {H.~A.}\ \bibnamefont {Padmore}},\ }\bibfield  {title} {\enquote {\bibinfo {title} {A novel system for measurement of the transverse electron momentum distribution from photocathodes},}\ }\href {\doibase 10.1063/1.4904930} {\bibfield  {journal} {\bibinfo  {journal} {Review of Scientific Instruments}\ }\textbf {\bibinfo {volume} {86}},\ \bibinfo {pages} {015103} (\bibinfo {year} {2015})}\BibitemShut {NoStop}%
\bibitem [{\citenamefont {Jones}\ \emph {et~al.}(2022)\citenamefont {Jones}, \citenamefont {Juarez-Lopez}, \citenamefont {Scheibler}, \citenamefont {Terekhov}, \citenamefont {Militsyn}, \citenamefont {Welsch},\ and\ \citenamefont {Noakes}}]{jones2022}%
  \BibitemOpen
  \bibfield  {author} {\bibinfo {author} {\bibfnamefont {L.}~\bibnamefont {Jones}}, \bibinfo {author} {\bibfnamefont {D.}~\bibnamefont {Juarez-Lopez}}, \bibinfo {author} {\bibfnamefont {H.}~\bibnamefont {Scheibler}}, \bibinfo {author} {\bibfnamefont {A.}~\bibnamefont {Terekhov}}, \bibinfo {author} {\bibfnamefont {B.}~\bibnamefont {Militsyn}}, \bibinfo {author} {\bibfnamefont {C.}~\bibnamefont {Welsch}}, \ and\ \bibinfo {author} {\bibfnamefont {T.}~\bibnamefont {Noakes}},\ }\bibfield  {title} {\enquote {\bibinfo {title} {{The measurement of photocathode transverse energy distribution curves (TEDCs) using the transverse energy spread spectrometer (TESS) experimental system}},}\ }\href {\doibase 10.1063/5.0109053} {\bibfield  {journal} {\bibinfo  {journal} {Review of Scientific Instruments}\ }\textbf {\bibinfo {volume} {93}},\ \bibinfo {pages} {113314} (\bibinfo {year} {2022})}\BibitemShut {NoStop}%
\bibitem [{\citenamefont {Sertore}\ \emph {et~al.}(2022)\citenamefont {Sertore}, \citenamefont {Bertucci}, \citenamefont {Bosotti}, \citenamefont {Giove}, \citenamefont {Monaco},\ and\ \citenamefont {Paparella}}]{sertore2022}%
  \BibitemOpen
  \bibfield  {author} {\bibinfo {author} {\bibfnamefont {D.}~\bibnamefont {Sertore}}, \bibinfo {author} {\bibfnamefont {M.}~\bibnamefont {Bertucci}}, \bibinfo {author} {\bibfnamefont {A.}~\bibnamefont {Bosotti}}, \bibinfo {author} {\bibfnamefont {D.}~\bibnamefont {Giove}}, \bibinfo {author} {\bibfnamefont {L.}~\bibnamefont {Monaco}}, \ and\ \bibinfo {author} {\bibfnamefont {R.}~\bibnamefont {Paparella}},\ }\bibfield  {title} {\enquote {\bibinfo {title} {{Assembly and Characterization of Low-Energy Electron Transverse Momentum Measurement Device (TRAMM) at INFN LASA}},}\ }in\ \href {\doibase 10.18429/JACoW-IPAC2022-THPOPT026} {\emph {\bibinfo {booktitle} {Proc. IPAC'22}}}\ (\bibinfo {year} {2022})\BibitemShut {NoStop}%
\bibitem [{\citenamefont {Abo-Bakr}\ \emph {et~al.}(2020)\citenamefont {Abo-Bakr}, \citenamefont {Al-Saokal}, \citenamefont {Anders}, \citenamefont {Bergmann}, \citenamefont {Bürkmann-Gehrlein} \emph {et~al.}}]{abo-bakr2020}%
  \BibitemOpen
  \bibfield  {author} {\bibinfo {author} {\bibfnamefont {M.}~\bibnamefont {Abo-Bakr}}, \bibinfo {author} {\bibfnamefont {N.}~\bibnamefont {Al-Saokal}}, \bibinfo {author} {\bibfnamefont {W.}~\bibnamefont {Anders}}, \bibinfo {author} {\bibfnamefont {Y.}~\bibnamefont {Bergmann}}, \bibinfo {author} {\bibfnamefont {K.}~\bibnamefont {Bürkmann-Gehrlein}},  \emph {et~al.},\ }\bibfield  {title} {\enquote {\bibinfo {title} {The {Berlin} {Energy} {Recovery} {Linac} {Project} {BERLinPro} - {Status}, {Plans} and {Future} {Opportunities}},}\ }in\ \href {\doibase 10.18429/JACOW-ERL2019-MOCOXBS04} {\emph {\bibinfo {booktitle} {Proc. ERL'19}}}\ (\bibinfo {year} {2020})\BibitemShut {NoStop}%
\bibitem [{\citenamefont {Kamps}\ \emph {et~al.}(2023)\citenamefont {Kamps}, \citenamefont {Abo-Bakr}, \citenamefont {Alberdi}, \citenamefont {Birke}, \citenamefont {Echevarria}, \citenamefont {Ergenlik}, \citenamefont {Fleischhauer}, \citenamefont {Frahm}, \citenamefont {Huck}, \citenamefont {Klemz},\ and\ \citenamefont {{others}}}]{kamps2022}%
  \BibitemOpen
  \bibfield  {author} {\bibinfo {author} {\bibfnamefont {T.}~\bibnamefont {Kamps}}, \bibinfo {author} {\bibfnamefont {M.}~\bibnamefont {Abo-Bakr}}, \bibinfo {author} {\bibfnamefont {B.}~\bibnamefont {Alberdi}}, \bibinfo {author} {\bibfnamefont {T.}~\bibnamefont {Birke}}, \bibinfo {author} {\bibfnamefont {P.}~\bibnamefont {Echevarria}}, \bibinfo {author} {\bibfnamefont {E.}~\bibnamefont {Ergenlik}}, \bibinfo {author} {\bibfnamefont {R.}~\bibnamefont {Fleischhauer}}, \bibinfo {author} {\bibfnamefont {A.}~\bibnamefont {Frahm}}, \bibinfo {author} {\bibfnamefont {H.}~\bibnamefont {Huck}}, \bibinfo {author} {\bibfnamefont {G.}~\bibnamefont {Klemz}}, \ and\ \bibinfo {author} {\bibnamefont {{others}}},\ }\bibfield  {title} {\enquote {\bibinfo {title} {Accelerator physics experiments at the versatile {SRF} photoinjector of {SEALab}},}\ }in\ \href {\doibase 10.18429/JACoW-IPAC2023-TUPL160} {\emph {\bibinfo {booktitle} {Proc. IPAC'23}}}\ (\bibinfo {year} {2023})\ pp.\ \bibinfo {pages} {2115--2118}\BibitemShut {NoStop}%
\bibitem [{\citenamefont {Kühn}\ \emph {et~al.}(2019)\citenamefont {Kühn}, \citenamefont {Al-Saokal}, \citenamefont {Bürger}, \citenamefont {Dirsat}, \citenamefont {Frahm}, \citenamefont {Jankowiak}, \citenamefont {Kamps}, \citenamefont {Klemz}, \citenamefont {Mistry}, \citenamefont {Neumann},\ and\ \citenamefont {Plötz}}]{kuhn2019}%
  \BibitemOpen
  \bibfield  {author} {\bibinfo {author} {\bibfnamefont {J.}~\bibnamefont {Kühn}}, \bibinfo {author} {\bibfnamefont {N.}~\bibnamefont {Al-Saokal}}, \bibinfo {author} {\bibfnamefont {M.}~\bibnamefont {Bürger}}, \bibinfo {author} {\bibfnamefont {M.}~\bibnamefont {Dirsat}}, \bibinfo {author} {\bibfnamefont {A.}~\bibnamefont {Frahm}}, \bibinfo {author} {\bibfnamefont {A.}~\bibnamefont {Jankowiak}}, \bibinfo {author} {\bibfnamefont {T.}~\bibnamefont {Kamps}}, \bibinfo {author} {\bibfnamefont {G.}~\bibnamefont {Klemz}}, \bibinfo {author} {\bibfnamefont {S.}~\bibnamefont {Mistry}}, \bibinfo {author} {\bibfnamefont {A.}~\bibnamefont {Neumann}}, \ and\ \bibinfo {author} {\bibfnamefont {H.}~\bibnamefont {Plötz}},\ }\bibfield  {title} {\enquote {\bibinfo {title} {Thermal {Load} {Studies} on the {Photocathode} {Insert} with {Exchangeable} {Plug} for the {bERLinPro} {SRF}-{Photoinjector}},}\ }in\ \href {\doibase 10.18429/JACOW-SRF2019-TUP100} {\emph {\bibinfo {booktitle} {Proc. SRF'19}}}\ (\bibinfo {year}
  {2019})\BibitemShut {NoStop}%
\bibitem [{\citenamefont {Kühn}\ \emph {et~al.}(2017)\citenamefont {Kühn}, \citenamefont {Borninkhof}, \citenamefont {Bürger}, \citenamefont {Frahm}, \citenamefont {Jankowiak}, \citenamefont {Kamps}, \citenamefont {Schmeißer}, \citenamefont {Schuster}, \citenamefont {Murcek}, \citenamefont {Teichert},\ and\ \citenamefont {Xiang}}]{kuhn2017}%
  \BibitemOpen
  \bibfield  {author} {\bibinfo {author} {\bibfnamefont {J.}~\bibnamefont {Kühn}}, \bibinfo {author} {\bibfnamefont {J.}~\bibnamefont {Borninkhof}}, \bibinfo {author} {\bibfnamefont {M.}~\bibnamefont {Bürger}}, \bibinfo {author} {\bibfnamefont {A.}~\bibnamefont {Frahm}}, \bibinfo {author} {\bibfnamefont {A.}~\bibnamefont {Jankowiak}}, \bibinfo {author} {\bibfnamefont {T.}~\bibnamefont {Kamps}}, \bibinfo {author} {\bibfnamefont {M.}~\bibnamefont {Schmeißer}}, \bibinfo {author} {\bibfnamefont {M.}~\bibnamefont {Schuster}}, \bibinfo {author} {\bibfnamefont {P.}~\bibnamefont {Murcek}}, \bibinfo {author} {\bibfnamefont {J.}~\bibnamefont {Teichert}}, \ and\ \bibinfo {author} {\bibfnamefont {R.}~\bibnamefont {Xiang}},\ }\bibfield  {title} {\enquote {\bibinfo {title} {{UHV} {Photocathode} {Plug} {Transfer} {Chain} {for} {the} {bERLinPro} {SRF}-{Photoinjector}},}\ }in\ \href {\doibase JACoW-IPAC2017-TUPAB029} {\emph {\bibinfo {booktitle} {Proc. {IPAC}'2017}}}\ (\bibinfo {year} {2017})\BibitemShut {NoStop}%
\bibitem [{\citenamefont {Spicer}(1958)}]{spicer1958}%
  \BibitemOpen
  \bibfield  {author} {\bibinfo {author} {\bibfnamefont {W.~E.}\ \bibnamefont {Spicer}},\ }\bibfield  {title} {\enquote {\bibinfo {title} {Photoemissive, photoconductive, and optical absorption studies of alkali-antimony compounds},}\ }\href {\doibase 10.1103/PhysRev.112.114} {\bibfield  {journal} {\bibinfo  {journal} {Physical Review}\ }\textbf {\bibinfo {volume} {112}},\ \bibinfo {pages} {114--114} (\bibinfo {year} {1958})}\BibitemShut {NoStop}%
\bibitem [{\citenamefont {Dowell}\ and\ \citenamefont {Schmerge}(2009)}]{dowell2009}%
  \BibitemOpen
  \bibfield  {author} {\bibinfo {author} {\bibfnamefont {D.}~\bibnamefont {Dowell}}\ and\ \bibinfo {author} {\bibfnamefont {J.~F.}\ \bibnamefont {Schmerge}},\ }\bibfield  {title} {\enquote {\bibinfo {title} {Quantum efficiency and thermal emittance of metal photocathodes},}\ }\href {\doibase 10.1103/PhysRevSTAB.12.074201} {\bibfield  {journal} {\bibinfo  {journal} {Physical Review Special Topics-Accelerators and Beams}\ }\textbf {\bibinfo {volume} {12}},\ \bibinfo {pages} {074201--074201} (\bibinfo {year} {2009})}\BibitemShut {NoStop}%
\bibitem [{\citenamefont {{Dassault Systems}}(2024)}]{CST}%
  \BibitemOpen
  \bibfield  {author} {\bibinfo {author} {\bibnamefont {{Dassault Systems}}},\ }\href@noop {} {\enquote {\bibinfo {title} {{CST Studio Suite - 3Dexperience}},}\ } (\bibinfo {year} {2024}),\ \bibinfo {note} {2024}\BibitemShut {NoStop}%
\bibitem [{\citenamefont {Flottmann}, \citenamefont {Lidia},\ and\ \citenamefont {Piot}(2003)}]{flottmann2003}%
  \BibitemOpen
  \bibfield  {author} {\bibinfo {author} {\bibfnamefont {K.}~\bibnamefont {Flottmann}}, \bibinfo {author} {\bibfnamefont {S.}~\bibnamefont {Lidia}}, \ and\ \bibinfo {author} {\bibfnamefont {P.}~\bibnamefont {Piot}},\ }\href@noop {} {\enquote {\bibinfo {title} {{Recent improvements to the ASTRA particle tracking code}},}\ }\bibinfo {type} {Tech. Rep.}\ (\bibinfo  {institution} {Lawrence Berkeley National Lab.(LBNL), Berkeley, CA (United States)},\ \bibinfo {year} {2003})\BibitemShut {NoStop}%
\bibitem [{\citenamefont {Kruschwitz}(2018)}]{Kruschwitz2018}%
  \BibitemOpen
  \bibfield  {author} {\bibinfo {author} {\bibfnamefont {J.}~\bibnamefont {Kruschwitz}},\ }\href {\doibase 10.1117/3.2500912} {\emph {\bibinfo {title} {Field Guide to Colorimetry and Fundamental Color Modeling}}},\ edited by\ \bibinfo {editor} {\bibfnamefont {J.}~\bibnamefont {Greivenkamp}},\ Vol.\ \bibinfo {volume} {FG42}\ (\bibinfo  {publisher} {SPIE Press},\ \bibinfo {year} {2018})\BibitemShut {NoStop}%
\bibitem [{\citenamefont {{CIE 2022}}(2022)}]{CIE1931_D65}%
  \BibitemOpen
  \bibfield  {author} {\bibinfo {author} {\bibnamefont {{CIE 2022}}},\ }\href@noop {} {\enquote {\bibinfo {title} {{CIE standard illuminant D65}},}\ } (\bibinfo {year} {2022}),\ \bibinfo {note} {international Commission on Illumination (CIE), Vienna, Austria, \url{https://doi.org/10.25039/CIE.DS.xvudnb9b}}\BibitemShut {NoStop}%
\bibitem [{\citenamefont {{CIE 2018}}(2018)}]{CIE1931_cmf}%
  \BibitemOpen
  \bibfield  {author} {\bibinfo {author} {\bibnamefont {{CIE 2018}}},\ }\href@noop {} {\enquote {\bibinfo {title} {{CIE 1931 colour-matching functions, 2 degree observer (data table)}},}\ } (\bibinfo {year} {2018}),\ \bibinfo {note} {international Commission on Illumination (CIE), Vienna, Austria, \url{https://doi.org/10.25039/CIE.DS.xvudnb9b}}\BibitemShut {NoStop}%
\bibitem [{\citenamefont {Davisson}\ and\ \citenamefont {Calbick}(1932)}]{Davisson1932}%
  \BibitemOpen
  \bibfield  {author} {\bibinfo {author} {\bibfnamefont {C.~J.}\ \bibnamefont {Davisson}}\ and\ \bibinfo {author} {\bibfnamefont {C.~J.}\ \bibnamefont {Calbick}},\ }\bibfield  {title} {\enquote {\bibinfo {title} {Electron lenses},}\ }\href {\doibase 10.1103/PhysRev.42.580} {\bibfield  {journal} {\bibinfo  {journal} {Phys. Rev.}\ }\textbf {\bibinfo {volume} {42}},\ \bibinfo {pages} {580--580} (\bibinfo {year} {1932})}\BibitemShut {NoStop}%
\bibitem [{\citenamefont {Yu}\ \emph {et~al.}(2021)\citenamefont {Yu}, \citenamefont {Wan}, \citenamefont {Tang},\ and\ \citenamefont {Feng}}]{Yu2021}%
  \BibitemOpen
  \bibfield  {author} {\bibinfo {author} {\bibfnamefont {L.}~\bibnamefont {Yu}}, \bibinfo {author} {\bibfnamefont {W.}~\bibnamefont {Wan}}, \bibinfo {author} {\bibfnamefont {W.-X.}\ \bibnamefont {Tang}}, \ and\ \bibinfo {author} {\bibfnamefont {J.}~\bibnamefont {Feng}},\ }\bibfield  {title} {\enquote {\bibinfo {title} {{Systematic analysis of a compact setup to measure the photoemitted electron beam transverse momentum and emittance}},}\ }\href {\doibase 10.1063/5.0013122} {\bibfield  {journal} {\bibinfo  {journal} {Review of Scientific Instruments}\ }\textbf {\bibinfo {volume} {92}},\ \bibinfo {pages} {013302} (\bibinfo {year} {2021})}\BibitemShut {NoStop}%
\bibitem [{\citenamefont {Maxson}\ \emph {et~al.}(2015)\citenamefont {Maxson}, \citenamefont {Cultrera}, \citenamefont {Gulliford},\ and\ \citenamefont {Bazarov}}]{maxson2015}%
  \BibitemOpen
  \bibfield  {author} {\bibinfo {author} {\bibfnamefont {J.}~\bibnamefont {Maxson}}, \bibinfo {author} {\bibfnamefont {L.}~\bibnamefont {Cultrera}}, \bibinfo {author} {\bibfnamefont {C.}~\bibnamefont {Gulliford}}, \ and\ \bibinfo {author} {\bibfnamefont {I.}~\bibnamefont {Bazarov}},\ }\bibfield  {title} {\enquote {\bibinfo {title} {{Measurement of the tradeoff between intrinsic emittance and quantum efficiency from a NaKSb photocathode near threshold}},}\ }\href {\doibase 10.1063/1.4922146} {\bibfield  {journal} {\bibinfo  {journal} {Applied Physics Letters}\ }\textbf {\bibinfo {volume} {106}},\ \bibinfo {pages} {234102} (\bibinfo {year} {2015})}\BibitemShut {NoStop}%
\bibitem [{\citenamefont {Kachwala}\ \emph {et~al.}(2023)\citenamefont {Kachwala}, \citenamefont {Saha}, \citenamefont {Bhattacharyya}, \citenamefont {Montgomery}, \citenamefont {Chubenko},\ and\ \citenamefont {Karkare}}]{kachwala2023}%
  \BibitemOpen
  \bibfield  {author} {\bibinfo {author} {\bibfnamefont {A.}~\bibnamefont {Kachwala}}, \bibinfo {author} {\bibfnamefont {P.}~\bibnamefont {Saha}}, \bibinfo {author} {\bibfnamefont {P.}~\bibnamefont {Bhattacharyya}}, \bibinfo {author} {\bibfnamefont {E.}~\bibnamefont {Montgomery}}, \bibinfo {author} {\bibfnamefont {O.}~\bibnamefont {Chubenko}}, \ and\ \bibinfo {author} {\bibfnamefont {S.}~\bibnamefont {Karkare}},\ }\bibfield  {title} {\enquote {\bibinfo {title} {Demonstration of thermal limit mean transverse energy from cesium antimonide photocathodes},}\ }\href {\doibase 10.1063/5.0159924} {\bibfield  {journal} {\bibinfo  {journal} {Applied Physics Letters}\ }\textbf {\bibinfo {volume} {123}},\ \bibinfo {pages} {044106} (\bibinfo {year} {2023})}\BibitemShut {NoStop}%
\end{thebibliography}%

\end{document}